\documentclass[]{aa}
\usepackage{graphicx}
\usepackage{epsfig}
\usepackage{subfigure}
\usepackage{amssymb}
\usepackage{amsmath}
\usepackage{textgreek}

\usepackage{txfonts}
\bibpunct{(}{)}{;}{a}{}{,}

\newcommand{\FIG}[1]{#1}

 \def\mso{\,{\rm M}_\odot}

 \def\lso{\,{\rm L}_\odot}

 \def\kms{\, {\rm km}\, {\rm s}^{-1}}
 
 \def\simle{\mathrel{\hbox{\rlap{\hbox{\lower4pt\hbox{$\sim$}}}\hbox{$<$}}}}
 \def\simgr{\mathrel{\hbox{\rlap{\hbox{\lower4pt\hbox{$\sim$}}}\hbox{$>$}}}}
 \def\vinf{\, \mathrm{v}_\infty}
 \def\mdot{\, \dot{M}}
 \def\msoy{\, \mso~{\rm yr}^{-1}}
 \def\muG{\,\textmu G}

\def\rb{\, R_{\rm B}}
\def\rd{\, R_{\rm D}}

\def\vel{\, \mathbf v}
\def\Bfield{\, \mathbf B}
\def\micron{\, \textmu m}
\def\UCam{\,\object{U\,Cam}}
\def\STCam{\,\object{ST\,Cam}}
\def\VYUma{\,\object{VY\,UMa}}

\begin{document}

   \title{Eyes in the sky}
   \subtitle{Interactions between AGB winds and the interstellar magnetic field
      \thanks{{\it Herschel} is an ESA space observatory with science instruments 
              provided by European-led Principal Investigator consortia and with important participation from NASA.}}
   \author{A.\,J. van Marle
          \inst{1}
          \and
          N.\,L.\,J. Cox
          \inst{1}
          \and
          L. Decin
          \inst{1,2}
          }

   \offprints{A. J. van Marle}

   \institute{KU Leuven, Institute of Astronomy, 
              Celestijnenlaan 200D, B-3001 Leuven, Belgium \\
              \email{AllardJan.vanMarle@ster.kuleuven.be}
	      \and Universiteit van Amsterdam, Sterrenkundig Instituut Anton Pannekoek, 
	           Science Park 904, NL-1098, Amsterdam, The Netherlands
}

\date{Received <date> / Accepted <date>}

\abstract
{The extended circumstellar envelopes (CSE) of evolved low-mass stars display a large variety of morphologies. 
Understanding the various mechanisms that give rise to these extended structures is important to trace their mass-loss history.}
{Here we aim to examine the role of the interstellar magnetic field in shaping the extended morphologies of slow dusty winds of Asymptotic Giant-branch (AGB)
stars in an effort to pin-point the origin of so-called \emph{eye} shaped CSE of three carbon-rich AGB stars. 
In addition, we seek to understand if this pre-planetary nebula (PN) shaping can be responsible for asymmetries observed in PNe.}
{Hydrodynamical simulations are used to study the effect of typical interstellar magnetic fields on the free-expanding spherical stellar winds as they sweep up the local interstellar medium (ISM).} 
{The simulations show that typical Galactic interstellar magnetic fields of 5 to 10\muG\, are sufficient to alter the spherical expanding shells of 
AGB stars to appear as the characteristic \emph{eye} shape revealed by far-infrared observations. 
The typical sizes of the simulated \emph{eyes} are in accordance with the observed physical sizes. 
However, the \emph{eye} shapes are of transient nature. 
Depending on the stellar and interstellar conditions they develop after 20\,000 to 200\,000~yrs and last for about 50\,000 to 500\,000~yrs, 
assuming that the star is at rest relative to the local interstellar medium. 
Once formed the eye shape will develop lateral outflows parallel to the magnetic field. 
The ``explosion'' of a PN in the center of the \emph{eye}-shaped dust shell gives rise to an asymmetrical nebula with prominent inward pointing Rayleigh-Taylor instabilities.}
{Interstellar magnetic fields can clearly affect the shaping of wind-ISM interaction shells. 
The occurrence of the \emph{eyes} is most strongly influenced by stellar space motion and ISM density. 
Observability of this transient phase is favoured for lines-of-sight perpendicular to the interstellar magnetic field direction. 
The simulations indicate that shaping of the pre-PN envelope can strongly affect the shape and size of PNe.}

  \titlerunning{Eyes in the sky}
  \authorrunning{van Marle, Cox \& Decin}


  \maketitle

%

\section{Introduction}
Asymptotic Giant Branch stars (AGB stars) occur in the late stages of the evolution of low mass stars \citep[][and references therein]{HabingOlofsson:2003,Herwig:2005}. 
AGB stars have a high luminosity ($\sim$1000-10\,000$\lso$) and low surface temperature ($\sim$2000-3500\,K) \citep[][and references therein]{HabingOlofsson:2003,Herwig:2005} and a slow, 
dust-driven wind \citep[e.g.][]{Kwok:1975,Lamerscassinelli:1999,ElitzurIvezi:2001}. 
Owing to their high mass loss rate ($10^{-7}-10^{-5}\msoy$) and the high dust content of their winds \citep[0.1-1\% according to ][]{Arndtetal:1997,Marengoetal:1997,Hoogzaadetal:2002,Woodsetal:2012}, 
AGB stars are considered the primary source of dust in the present-day interstellar medium (ISM) \citep[][and references therein]{HabingOlofsson:2003,Herwig:2005,Andersen:2007,Shukovskaetal:2008,ZhukovskaHenning:2013}. 
Because low mass stars do not have a strong wind until they reach the AGB phase, 
the AGB wind interacts directly with the ISM, rather than with the stellar wind remnants from earlier evolutionary stages, 
as would be the case for more massive stars \citep{Villaveretal:2002,Villaveretal:2012}. 
The collision between the stellar wind and the ISM results in the formation of 
a high density shell of swept-up interstellar gas, which is pushed outwards into the ISM. 

\emph{Herschel}/PACS (\citealt{Pilbrattetal:2010}, \citealt{Poglitschetal:2010}) observations obtained within the context of
the MESS programme (\citealt{Groenewegenetal:2011}) revealed a large variety of faint complex extended emission structures
around AGB stars and red supergiants (\citealt{Coxetal:2012}). Several types of morphological classes could be
discerned: 1) \emph{fermata} (bow waves), 2) \emph{eyes}, 3) \emph{rings}, 4) \emph{irregular}, and 5) point sources.
\emph{Fermata}-type structures are due to interaction of the slow dusty stellar wind of an evolved star with its surrounding
medium. Hydrodynamical simulations of these interactions have been able to predict and reproduce the observations in
great detail (e.g. \citealt{Villaveretal:2003}, \citealt{Wareingetal:2007}, \citealt{vanMarleetaldust:2011}, \citealt{Coxetal:2012}, 
\citealt{Decinetal:2012}, \citealt{vanMarleetal:2014}).
Ring or shell-type structures can be formed by the interaction of a spherically symmetric wind with a smooth ISM and 
are also expected from sudden changes in the mass-loss rate during the AGB phase 
\citep[e.g.][]{Villaveretal:2002,Villaveretal:2003,Kerschbaumetal:2010} created by e.g. a thermal pulse \citep{Maerckeretal:2005}.
Irregular structures of several objects in the MESS sample have also been studied in detail. 
In these cases the irregularity was seen to arise from additional mechanisms such as jets ({\object{R\,Aqr}) and spiral
structures (\object{W\,Aql}) from binary interaction \citep{Mayeretal:2013}.
Objects classified as a point source were on average more distant and the absence of extended emission could hence
represent the angular resolution and surface brightness limits.

\subsection{Eyes in the sky}
The more mysterious class constitutes several carbon-rich evolved stars whose resolved, detached emission is
axi-symmetric (rather than spherical symmetric) with an \emph{eye}-like shape. The typical distance from the \emph{pupil} to 
the \emph{eyelid} is $\sim$0.1-0.2~pc, with a minor/major axis ratio of $\sim$1.6.
Deeper follow-up PACS 100 and 160~\micron\, observations of the original data presented in \citet{Coxetal:2012} 
confirm the \emph{eye} shape for 3 stars, \object{\STCam}, \object{\UCam}, and \object{\VYUma}%
\footnote{\emph{Herschel} obsids: 1342242567,1342242568 (\STCam), 1342242577, 1342242578 (\UCam),
1342243219, 1342243220 (\VYUma), 1342241436, 1342241437 (\object{AQ\,Sqr}), 1342237178, 1342237179 (\object{UX\,Dra})}. 
The false-colour images composed of PACS 70\micron\, (blue), 100\micron\, (green), and 160\micron\, (red) images are shown in Fig.~\ref{fig:eyes}.
\object{UX\,Dra}, previously included in the \emph{eye} class, could now be re-classified as a likely \emph{ring}-type 
(c.q. detached shells), while for \object{AQ\,Sqr} the extended emission is now clearly of the \emph{fermata}-type. 
The classification of the remaining \emph{eye} objects (\object{R\,Crt}, \object{V\,Pav}) could not be confirmed and remains uncertain.

Basic stellar parameters for the three carbon-rich AGB stars with \emph{eye}-like extended emission,
\UCam, \STCam, and \VYUma\ are provided in Table~\ref{tab:targets}.
Terminal wind velocities and absolute space motion are accurate to within a few $\kms$.
Distances, and hence mass-loss rates, are less accurate. Literature values for distances range from 330 to 525~pc for \UCam,
from 390 to 800~pc for \STCam, and from 445 to 673~pc for \VYUma. For consistency we adopted the values for 
the distance, mass loss rate $(\dot{M})$, effective temperature (T$_\mathrm{eff}$), and wind velocity $(v_\infty)$ 
from a single reference, \citet{BergeatChevallier:2005}.

Some Planetary Nebulae (PNe), e.g. \object{M57}, \object{NG\,6543} and, especially, \object{MyCn18}, also exhibit \emph{eye}-like shapes. 
PNe are formed through the interaction between a fast post-AGB type wind and its slower AGB wind predecessor \citep{Kwoketal:1978} 
and particular shapes are usually attributed to density variations in the two winds \citep[][and references therein]{Kwok:2000}. 
However, such an explanation does not work for the circumstellar shells of AGB stars. 
The AGB wind collides directly with the surrounding ISM. 
Density variations in the ISM may deform the shell, but are unlikely to result in the sort of mirror symmetry, 
which we observe in the \emph{eye}-like structures. 
The influence of interstellar magnetic fields on the shape of planetary nebulae (PNe) was predicted analytically by \citet{Heiligman:1980,SokerDgani:1997} and
more recently \citet{FalcetaGoncalvesMonteiro:2014} showed numerical models of planetary nebulae and concluded that only 
strong magnetic fields ($\simeq\,500$\muG) could create a bipolar nebulae.

The influence of interstellar magnetic fields on the circumstellar medium was explored for large scale structures (e.g. superbubbles)
by \citet{Tomisaka:1990,Tomisaka:1992} and \citet{Ferriere:1991}. \citet{vanMarleetal:2014} 
showed the effect of an interstellar magnetic field on the bow bow shock of the red supergiant \object{\textalpha-Orionis}. 
In this paper we explore the possibility that the peculiar, \emph{eye}-like shape of the circumstellar shells around the AGB stars
\UCam, \STCam, and \VYUma, is the result of the interaction between the AGB wind and the interstellar magnetic field. 

\subsection{Layout}
In Sections~\ref{sec-analytic} and~\ref{sec-method} we introduce the analytical approximation and numerical method, respectively.
Sect.~\ref{sec-results} presents our results for the numerical simulations which cover a parameter space for different AGB mass loss 
rates, interstellar densities and magnetic field strengths. 
The validity of our 2.5-D approach is confirmed with a 3-D model simulation (Sect.~\ref{sec-3D}). 
We also demonstrate the effect of a warm ISM on the shape of th circumstellar shell (Sect.~\ref{sec-warmISM}).  
In Sect.~\ref{sec-discussion} we discuss the (likelihood of) formation of the \emph{eye}-shape in the context of AGB winds and the interaction
with the interstellar medium. 
Because PNe are formed inside the AGB-wind bubbles, it is possible that the shape of an AGB-wind bubble influences the 
evolution of the PN. 
Therefore, we briefly demonstrate the evolution of a PN inside the bubbles predicted by our simulations (Sect.~\ref{sec-PN}).

\begin{table*}[htp!]
\caption{Basic stellar parameters of the three carbon-rich \emph{eye} objects.}
\begin{tabular}{lcccccccccc}
\hline\hline
Object			& distance\tablefootmark{a}& l	&   b    	& $\mdot$\tablefootmark{a} 	& $\vinf$\tablefootmark{a} &    $v_\star$\tablefootmark{b}	& North-east (PA) & \multicolumn{2}{c}{Length (pc)\tablefootmark{c}}\\	
			& (pc)		& (\degr)	& (\degr) 	& $\msoy$ 	       		& ($\kms$)		   &	($\kms$)                        & (\degr)	& Semi-minor & Semi-major \\ \hline
U\,Cam			& 525        	&  141.1499 	&  +05.9679	& $2.0\times10^{-6}$  	 	& 20.6			   &	  9.8          			& 350.4		& 0.16	     	  & 0.28  \\
ST\,Cam			& 800           &  142.3898 	&  +14.9236	& $1.1\times10^{-6}$   		& 9.0			   &	 28.7       			& 247.8		& 0.30	     	  & 0.47  \\
VY\,UMa			& 445        	&  139.5898 	&  +45.4137	& $1.4\times10^{-7}$   		& 6.0			   &	 28.2           		& 70.5		& 0.10	       	  & 0.16   \\ 
\hline
\end{tabular}
\tablefoot{
\tablefoottext{a}{Distance, $\mdot$, and $\vinf$ adopted from \citet{BergeatChevallier:2005}.} \\
\tablefoottext{b}{Space velocities have been corrected for the motion of the sun [(U, V, W) = (11.10, 12.24, 7.25) $\kms$ ] \citep{Schronrichetal:2010}.}\\
\tablefoottext{c}{Minor and major axis of the \emph{eye} structures }}
\label{tab:targets}
\end{table*}

\begin{figure*}[hp!]
\FIG{
 \centering
\mbox{
\subfigure{\includegraphics[width=0.33\textwidth]{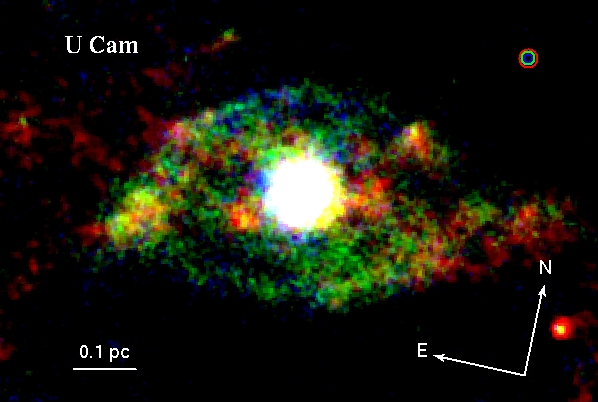}}
\subfigure{\includegraphics[width=0.33\textwidth]{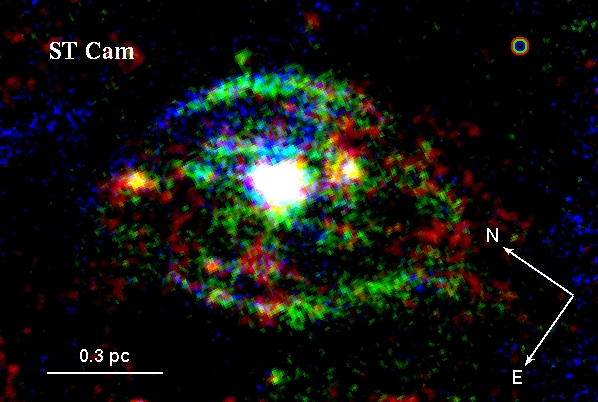}}
\subfigure{\includegraphics[width=0.33\textwidth]{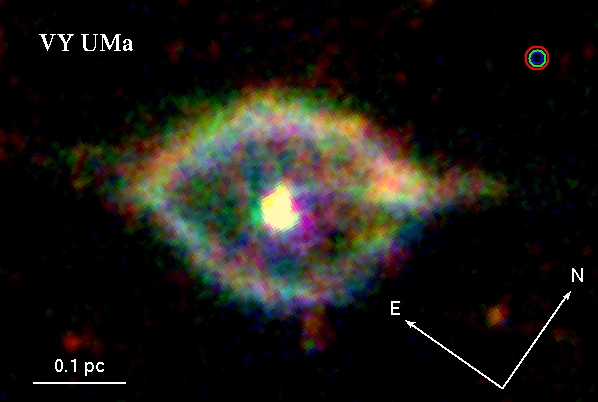}}}}
\caption{\emph{Herschel}/PACS 70~$\mu$m, 100~$\mu$m, 160~$\mu$m false-colour composite image of U\,Cam (left),
 ST\,Cam (middle), and VY\,Uma (right). Spatial scales are indicated with the horizontal bar. The compass is 
oriented with galactic coordinates and the arms are 1\arcmin\ in length (note that the images have been rotated
to align the symmetry axis for each object). The PACS beam sizes (FWHM)
at 70~$\mu$m, 100~$\mu$m, 160~$\mu$m, are represented by the small blue, green, and red circles in the upper right
corner of each panel.}
\label{fig:eyes}
\end{figure*}

\section{Analytical approximation}\label{sec-analytic}
The influence of the interstellar magnetic field on the shape of circumstellar bubbles was shown in 
simulations by \citet{Tomisaka:1990,Tomisaka:1992} and \citet{Ferriere:1991}. 
These models proved that even a weak interstellar field ($\sim$5\muG) could significantly alter the shape of the circumstellar bubble. 
Expansion perpendicular to the magnetic field is slowed down by the magnetic pressure, turning the originally spherical bubble into an ovoid shape.
The characteristic length scale at which this occurs can be determined by comparing the ram pressure of the stellar wind:
\begin{equation}
 P_{\rm ram}~=~\frac{\mdot\vinf}{4 \pi R^2}
\end{equation}
with the magnetic field pressure of the interstellar magnetic field:
\begin{equation}
 P_{\rm B}~=~\frac{B^2}{2 \mu_0}
\end{equation}
with, $\mdot$ the stellar mass-loss rate and $\vinf$ the stellar wind velocity, $R$ the distance from the star, 
$B$ the absolute magnetic field strength and $\mu_0$ the magnetic permeability of
the interstellar medium, which is $4\pi$ in cgs. units. 
Equalizing these two pressure terms gives us, 
\begin{equation}
\rb~=~\frac{1}{B}\sqrt{2\mdot \vinf},
\label{eq:rb}
\end{equation}
with $\rb$ the distance from the star at which the ram pressure of the wind becomes smaller than the interstellar magnetic field pressure. 
If the free-streaming wind expands to this distance, the magnetic field will block any further expansion of the wind in the direction perpendicular to the field. 
Therefore, comparing $\rb$ gives us a measure for the size a circumstellar bubble has to reach in order for the interstellar magnetic field to be come relevant. 

Using typical input parameters we find that for a solar type star ($\mdot\simeq10^{-14}\msoy$, $\vinf\simeq500\kms$) $\rb$ is approximately 100\,AU 
(see Fig.~\ref{fig:rb}), 
which is actually typical for the size of the heliosphere. 
For a massive star, O-type star ($\mdot\simeq10^{-7}\msoy$, $\vinf\simeq2000\kms$), $\rb$ will be of the order of 2-3 pc.
For an AGB wind, which has a low velocity ($\simeq$10-15$\kms$), and a relatively high mass loss rate ($10^{-7}-10^{-5}\msoy$), 
the stagnation point will typically lie at 0.5-1.5\,pc (see Fig.~\ref{fig:rb}). 
Assuming a wind velocity of $10\kms$, a typical AGB life-time of 1\,Myr, and no other obstruction to the expanding wind, the AGB wind 
would reach a maximum distance of 10\,pc from the star. 
Therefore, the interstellar magnetic field cannot be disregarded in models of the circumstellar bubbles of AGB stars. 
N.B. $\rb$ denotes the termination shock radius for which further expansion perpendicular to the field stops completely. 
Even before the wind reaches this point, the influence of the magnetic field will start to deform the spherical symmetry of the circumstellar bubble. 
   
\begin{figure}
\FIG{
 \centering
\mbox{
\subfigure{\includegraphics[width=\columnwidth]{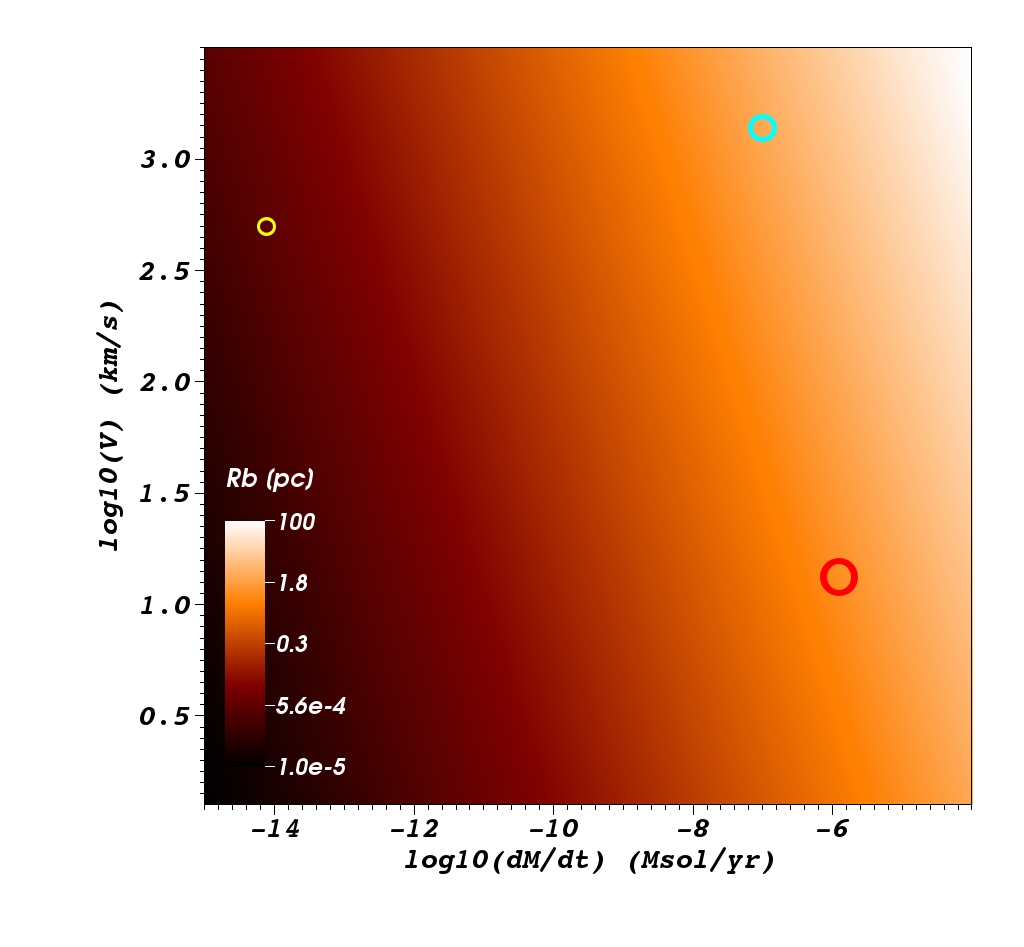}}}}
\caption{Magnetic-ram pressure stand-off distance for the wind termination shock as a function of mass loss rate and wind velocity, assuming 
an interstellar magnetic field strength of 10\muG. 
The three coloured circles give the location for a solar type star (yellow), an AGB star (red) and en O-star (blue)}
\label{fig:rb}
\end{figure}

   \begin{table}
   \centering
      \caption[]{Model input parameters.}
         \label{tab:Modpar}
         \resizebox{\columnwidth}{!}{
         \begin{tabular}{p{0.2\linewidth}ccccc}
            \hline \hline
            \noalign{\smallskip}
            Simulation      &  $B_{\rm ISM}$ & $n_{\rm ISM}$ & $\mdot$    & $\vinf$ 	    & Grid    \\
                            &  [\muG]        & [cm$^{-3}$]   & [$\msoy$]  & [km\,s$^{-1}$]  &         \\
            \noalign{\smallskip}
            \hline
            \noalign{\smallskip}
            A1              &   0           &  2               &  $10^{-7}$ & 10   & 2.5-D       \\
            A2              &   0           &  20              &  $10^{-6}$ & 10   & 2.5-D       \\
           
            \noalign{\smallskip}
            \hline
            \noalign{\smallskip}              
            B1               &   5           &  2              &  $10^{-7}$ & 10   & 2.5-D       \\
            B2               &   5           &  2              &  $10^{-6}$ & 10  & 2.5-D       \\
            B3               &   5           &  20             &  $10^{-7}$ & 10   & 2.5-D       \\
            B4               &   5           &  20             &  $10^{-6}$ & 10   & 2.5-D       \\            
            \noalign{\smallskip}
            \hline
            \noalign{\smallskip}            
            C1               &   10           &  2              &  $10^{-7}$ & 10  & 2.5-D       \\
            C2               &   10           &  2              &  $10^{-6}$ & 10  & 2.5-D       \\
            C3               &   10           &  20             &  $10^{-7}$ & 10  & 2.5-D       \\
            C4               &   10           &  20             &  $10^{-6}$ & 10  & 2.5-D      \\                                
            \noalign{\smallskip}
            \hline
            \noalign{\smallskip}            
            D1              &   5           &  2              &  $10^{-7}$ & 10   & 3-D, 	\\
			    &		    &		      & 	   &      & low resolution   \\          

            \noalign{\smallskip}
            \hline
            \noalign{\smallskip}            
            E1              &   5           &  2              &  $10^{-7}$ & 10   & 2.5-D, 	\\
			    &		    &		      & 	   &      & warm ISM   \\			  
	    \noalign{\smallskip}                       
            \hline
         \end{tabular}
         }
   \end{table}

\section{Numerical method}
\label{sec-method}
We use the {\tt MPI-AMRVAC} hydrodynamics code \citep{Keppensetal:2012}, 
which solves the conservation equations for mass, momentum and energy
\begin{eqnarray}
\frac{\partial \rho}{\partial t}+\nabla \cdot (\rho \vel)&=&0, \\
\rho \biggl(\frac{\partial \vel}{\partial t}~+~\vel\cdot\nabla\vel\biggr)~+~\nabla p_{\rm tot} -\frac{1}{\mu_0}(\nabla\times\Bfield)\times\Bfield~&=&~0, \\
\frac{\partial e}{\partial t} +\nabla \cdot (e {\vel} ) + \nabla \cdot (p\vel)+\biggl(\frac{\rho}{m_h}\biggr)^2 \Lambda(T)~&=&~0, 
\label{eq:mhd}
\end{eqnarray}
with $\rho$ the density, $\vel$ the velocity vector, $p_{\rm tot}$ the sum of the thermal pressure and the magnetic pressure, 
$\Bfield$ the magnetic field, 
and $m_h$ the hydrogen mass. 
The energy density $e$ is defined as the sum of thermal, kinetic and magnetic energy density,
\begin{equation}
e~=~\frac{p}{\gamma-1}~+~\frac{\rho v^2}{2}~+~\frac{B^2}{2\mu_0}, 
\end{equation}
with the adiabatic index $\gamma=5/3$, $p$ the thermal pressure and $v=|\vel|$ and $B=|\Bfield|$ the scale values of the velocity and the magnetic field respectively. 
We assume ideal MHD, which means that we have no local source terms for the magnetic field and do not account for plasma-reltaed physics.  
Therefore,
\begin{equation}
\frac{\partial \Bfield}{\partial t}~-~\nabla\times({\vel \times \Bfield})~=~0. 
\end{equation}
We keep the magnetic field divergence-free, using the method described by \citet{Powelletal:1999}.
The energy equation includes the effect of radiative cooling, which depends on local density, 
as well as a temperature dependent cooling curve for solar-metallicity gas, $\Lambda(T)$. 
We obtain $\Lambda(T)$ from \citet{Schureetal:2009}, which is a combination of a cooling curve generated with the {\tt SPEX} code \citep{KaastraMewe:2000} for temperatures above $10^4$\,K 
and the cooling curve from \citet{DalgarnoMcCray:1972} for lower temperatures.
\citet{Schureetal:2009} provide several curves for the lower temperature regime, depending on the 
ionization fraction. 
For our models we have chosen to assume that the ionization fraction is $10^{-3}$ below 10\,000\,K, 
 which is reasonable for the typical densities in our simulations \citep[e.g.][]{Shu:1983,Fatuzzoetal:2006}. 
Under these conditions, we can ignore the effect of ambipolar diffusion, which might cause a drift between 
the charged and neutral particles in the gas. 
For example, \citet{Casellietal:1998} found typical ambipolar diffusion time scales of 5\,Myr 
even for the much lower ionization fractions inside dense cores of molecular clouds, longer than the expected lifetime of an AGB star.  

We set a minimum temperature of 100\,K throughout the simulation, which limits the amount of compression due to radiative cooling, 
preventing numerical problems that can arise because of extreme compression of radiatively cooling gas. 

Our physical domain consist of a 2.5-D cylindrical grid in the $r,z$-plane, covering a physical space of 7.5 by 15\,pc. 
At the lowest level of resolution, we have 160$\times$320 grid points. 
The grid is allowed to be refined an additional four times, doubling the resolution at each new level of refinement 
for an effective resolution of 2560$\times$5120 grid points. 
This grid is filled with a constant density ISM with a temperature of 100\,K and a magnetic field that runs parallel with the $z$-axis. 
On the polar axis, we fill a small half-sphere ($R=0.05$\,pc) with stellar wind material. 
The part of the grid containing this half-sphere is always kept at the highest resolution. 
We assume the boundaries of the grid to be continuous, so that any signal that reaches the boundary will flow out of the grid. 
The only exception is the boundary condition at the polar axis ($R\,=\,0$), which we assume to be reflective.

\subsection{Input parameters}
The input parameters are specified in Table~\ref{tab:Modpar}. 
Simulations~A1 and A2 are the control models, which lack an interstellar magnetic field. 
Simulations~B1 through B4 show the interaction of the AGB wind with a 5\muG\,magnetic field, with varying interstellar densities and mass loss rates. 
Simulations~C1 through C4 repeat this for a stronger, 10\muG, magnetic field. 
Each simulation runs for 1\,Myr, which is typical of the timescales of an AGB-star lifetime \citep[][and references therein]{HabingOlofsson:2003,Herwig:2005}. 
For simulation~D1, we use the same input parameters as B1 in a low-resolution, 3-D model, for a period of 500\,000 years. 
{Finally, for simulation~E1, we use the same input parameters as B1, but with a warm ISM set at 8000\,K.}

\begin{figure*}
\FIG{
 \centering
\mbox{
\subfigure
{\includegraphics[width=0.5\textwidth]{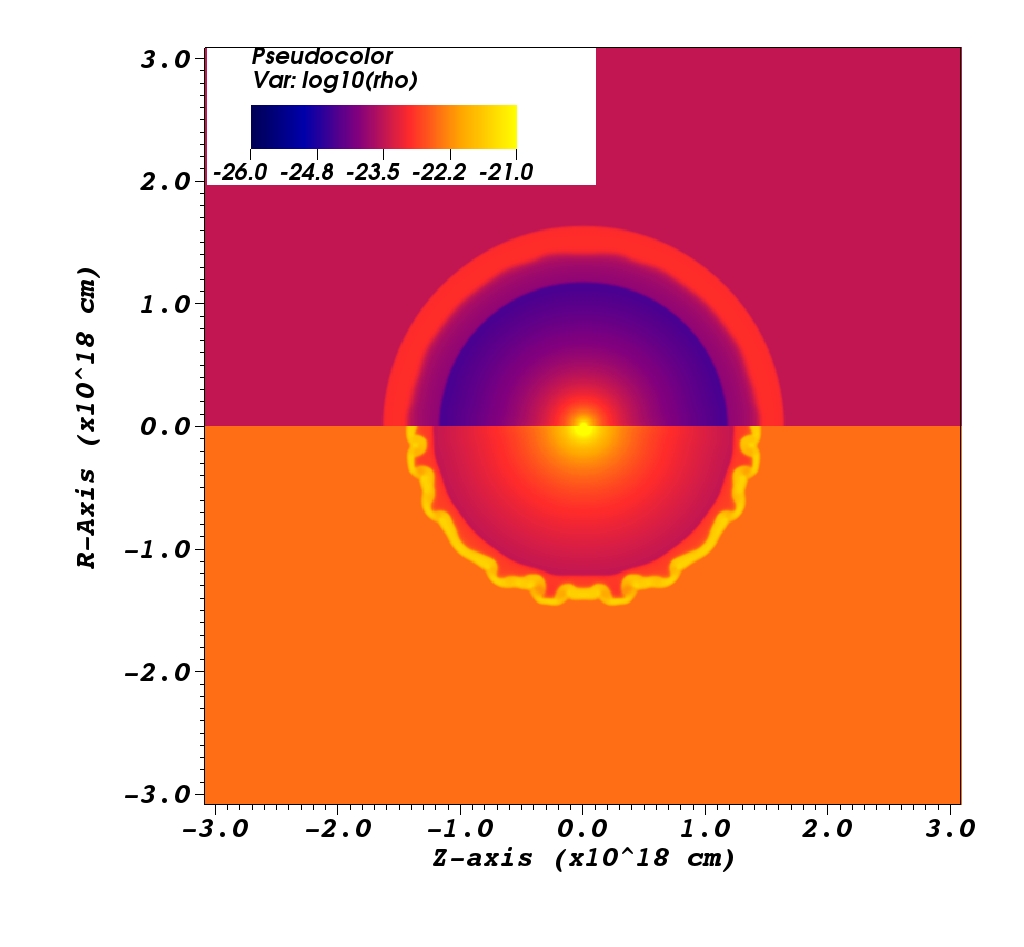}}
\subfigure
{\includegraphics[width=0.5\textwidth]{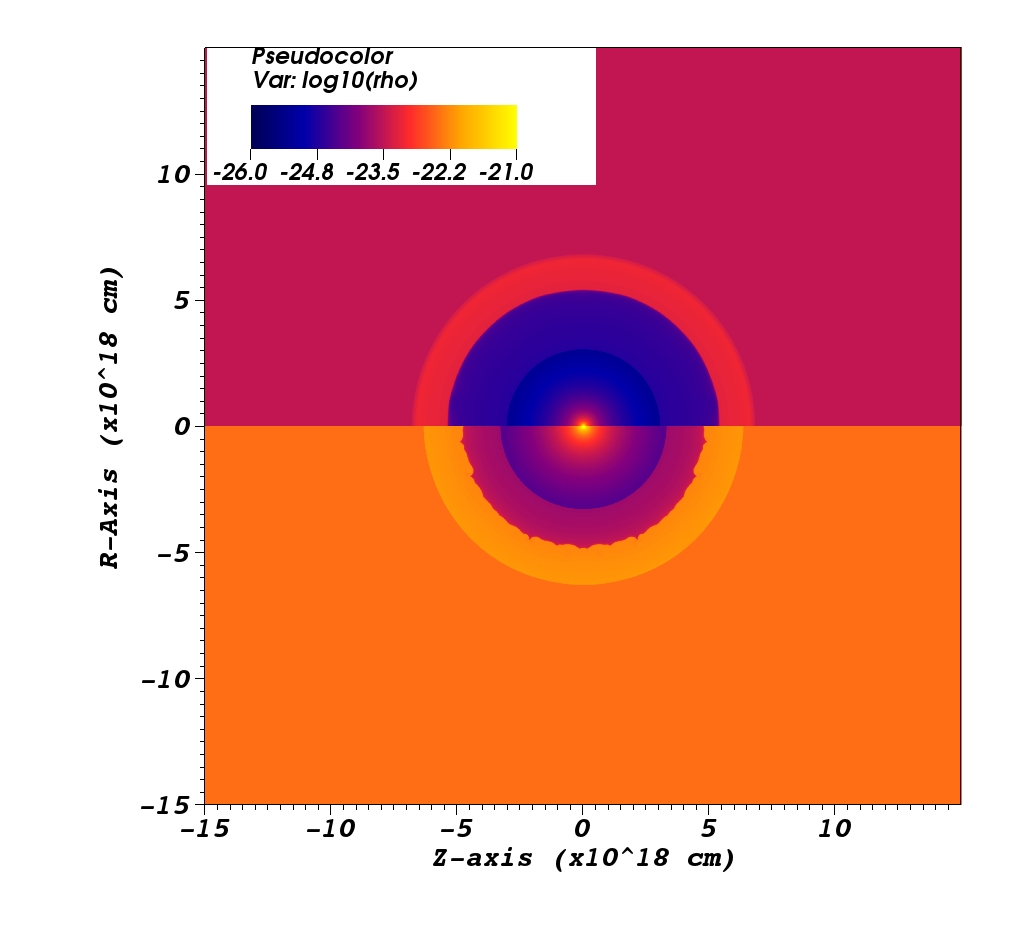}}}
}
\caption{Gas density of the interstellar medium for simulations A1 (top) and A2 (bottom) after 0.1 (left) and 1\,Myr (right). 
Physical sizes are 2$\times$2~pc (left panel) and 10$\times$10~pc (right panel).
In the earlier stages of the expansion the shells are compressed and simulation~A2 shows sign of thin-shell instabilities. 
Over time the interaction becomes more adiabatic; the compression of the gas diminishes; and the instabilities disappear.}
 \label{fig:A1A2}
\end{figure*}

\begin{figure*}
\FIG{
 \centering
\mbox{
\subfigure
{\includegraphics[width=0.5\textwidth]{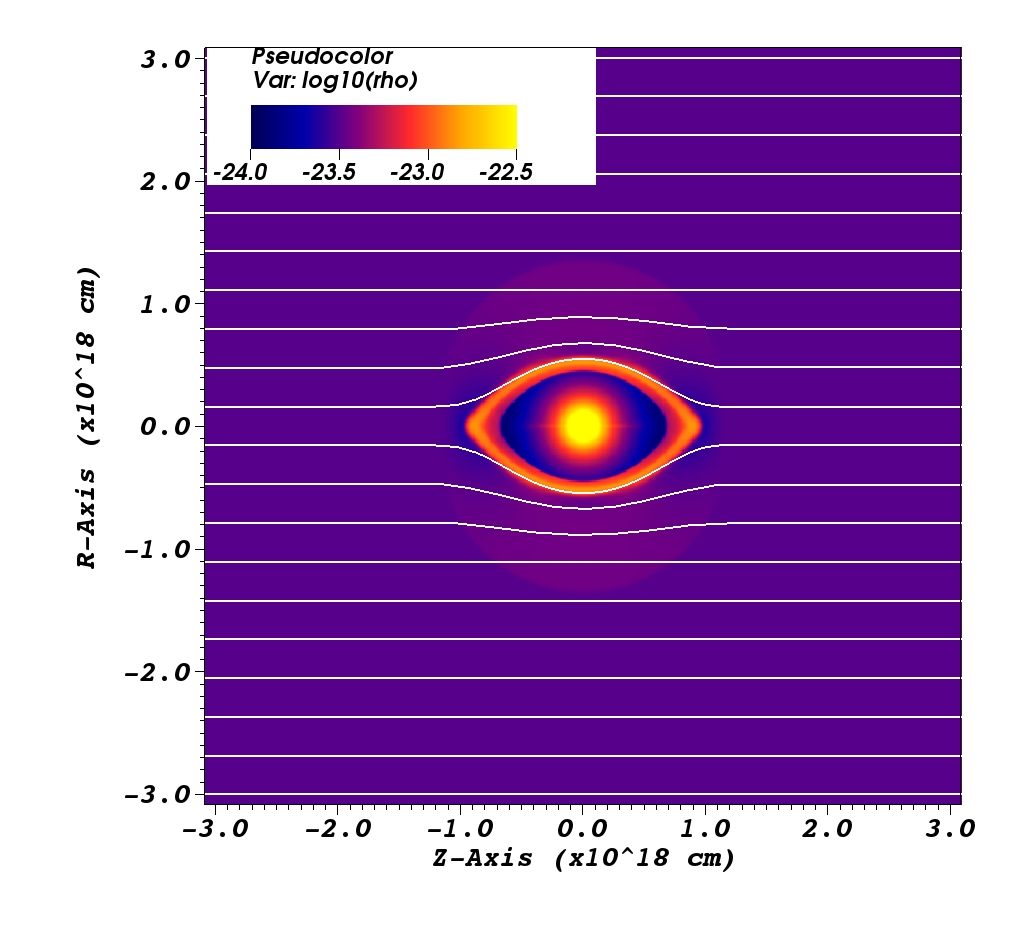}}
\subfigure
{\includegraphics[width=0.5\textwidth]{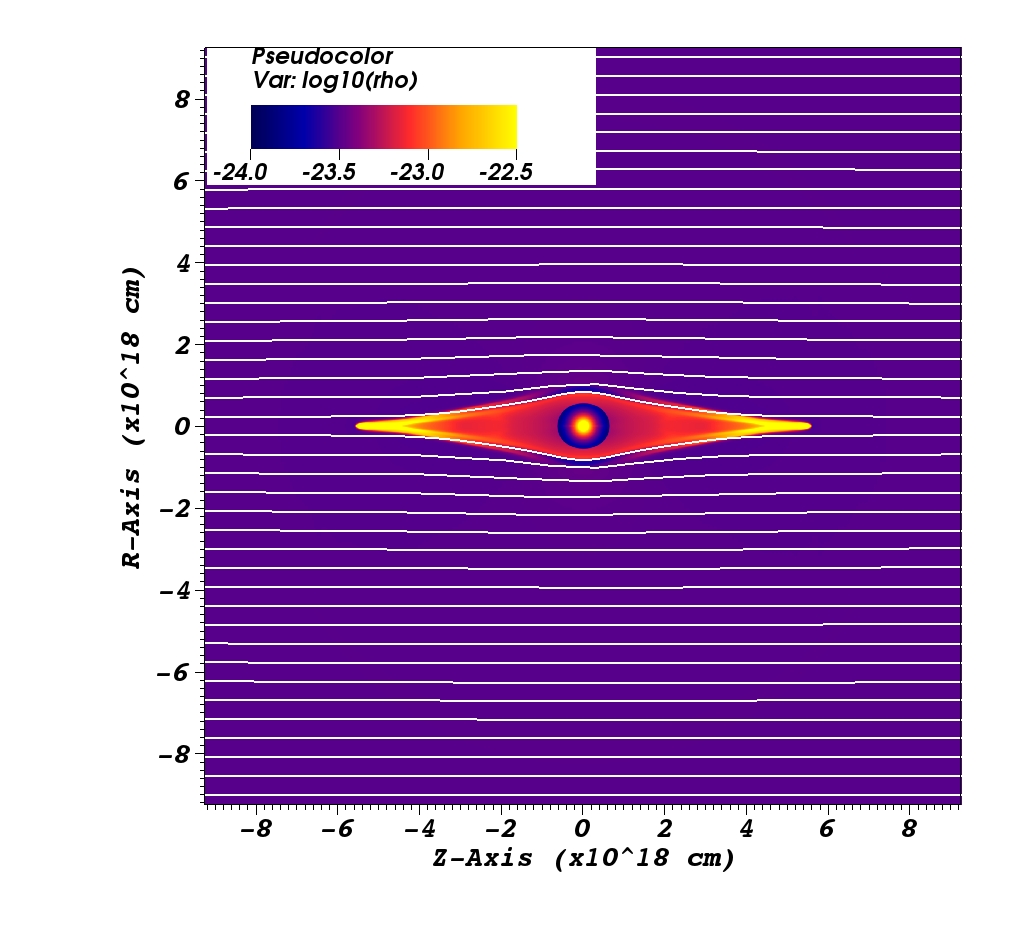}}}
}
\caption{Gas density and magnetic field lines for simulation~B1 after 40\,000 years (left, the onset of the \emph{eye} shape) 
and 300\,000 years (right, the beginning of the jet formation.)
Physical scales are 2$\times$2~pc (left panel) and 6$\times$6~pc (right panel).
}
 \label{fig:B1}
\end{figure*}

\begin{figure*}
\FIG{
 \centering
\mbox{
\subfigure
{\includegraphics[width=0.5\textwidth]{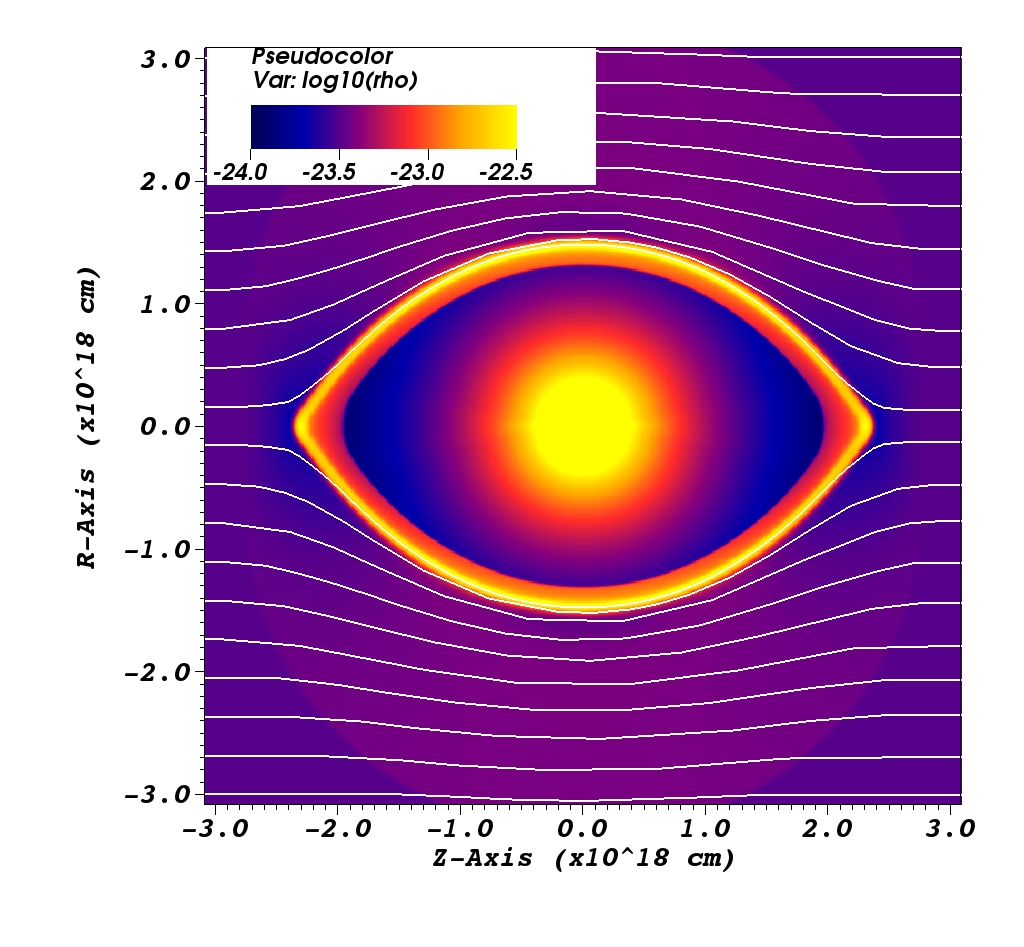}}
\subfigure
{\includegraphics[width=0.5\textwidth]{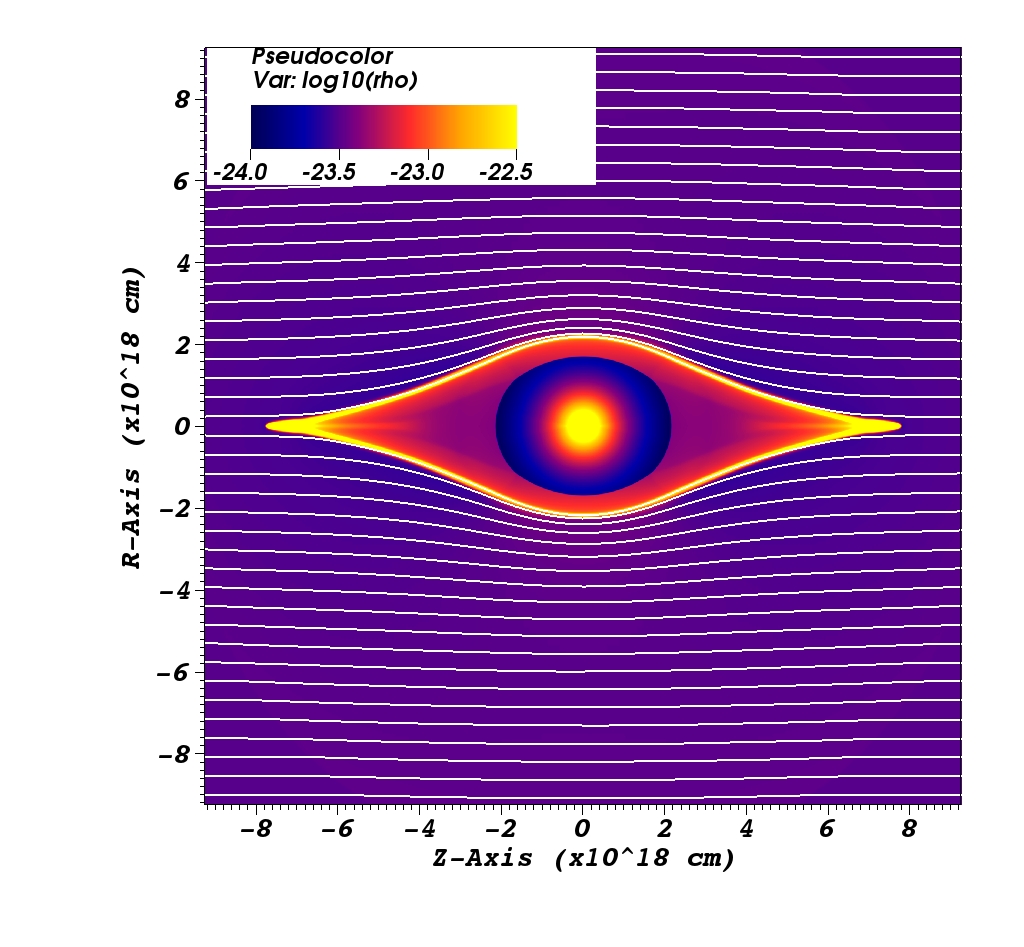}}}
}
\caption{Similar to Fig.~\ref{fig:B1}, but for simulation~B2. The \emph{eye} shape starts after 100\,000 years (left) and the collimated flows appear after 380\,000~years (right).
Physical scales are identical to those in Fig.~\ref{fig:B1}.
}
 \label{fig:B2}
\end{figure*}

\begin{figure*}
\FIG{
 \centering
\mbox{
\subfigure
{\includegraphics[width=0.5\textwidth]{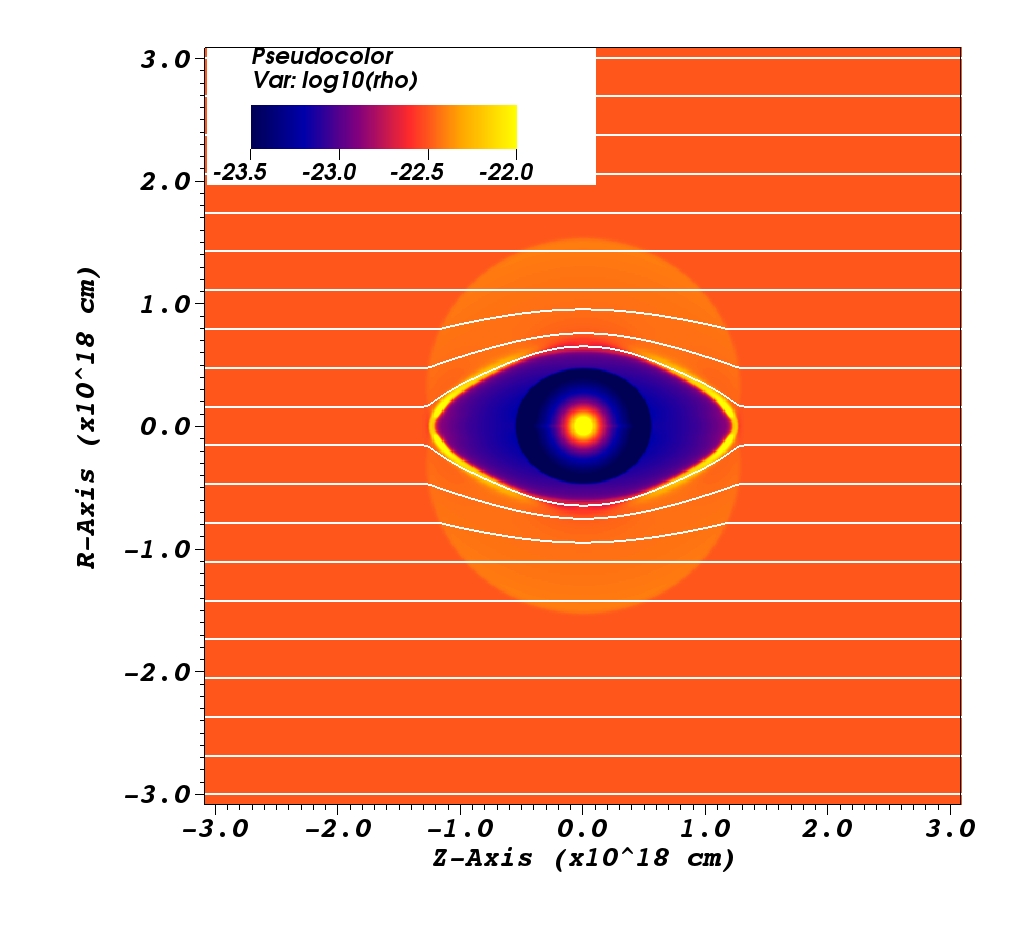}}
\subfigure
{\includegraphics[width=0.5\textwidth]{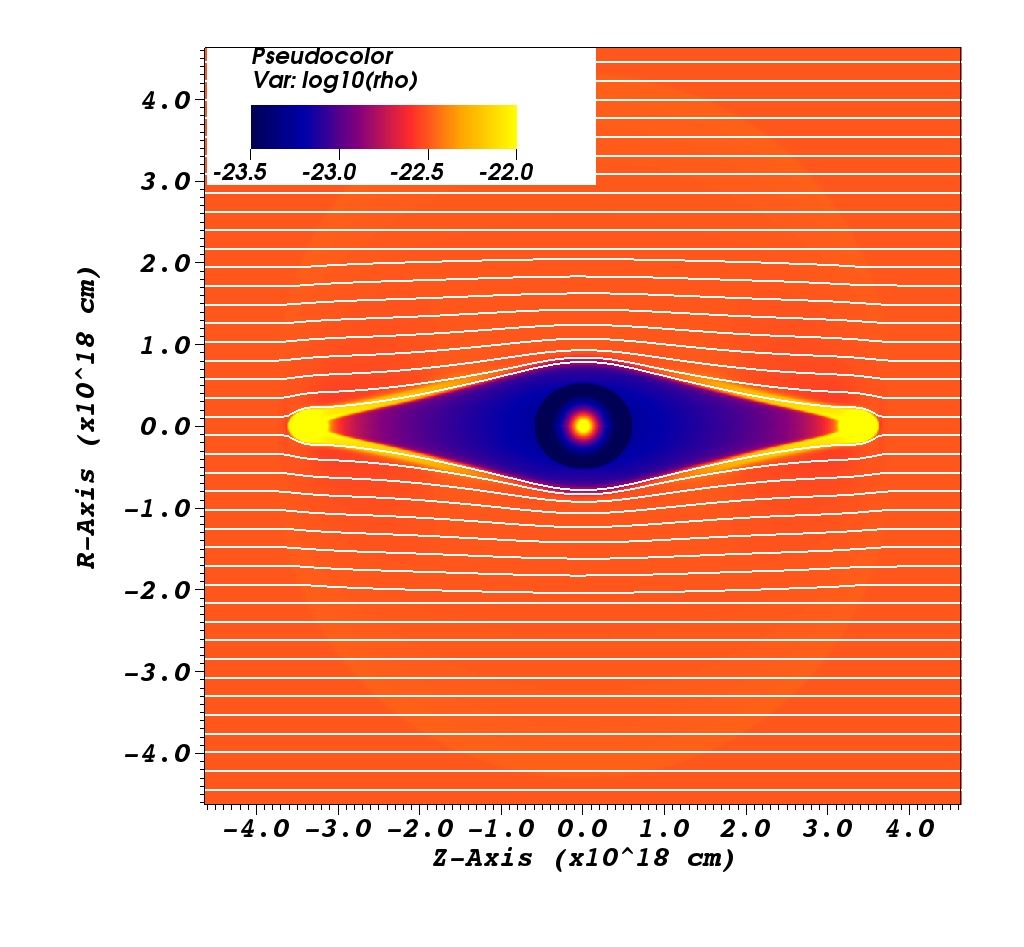}}}
}
\caption{Similar to Figs.~\ref{fig:B1}-\ref{fig:B2}, but for simulation~B3. The \emph{eye} shape starts after 120\,000 years (left) and the collimated flows appear after 430\,000~years (right).
Physical scales are 2$\times$2~pc (left panel) and 3$\times$3~pc (right panel). N.B. The scale of this figure deviates from the right panels of Figs.~\ref{fig:B1},\ref{fig:B2} and \ref{fig:B4}. 
}
 \label{fig:B3}
\end{figure*}

\begin{figure*}
\FIG{
 \centering
\mbox{
\subfigure
{\includegraphics[width=0.5\textwidth]{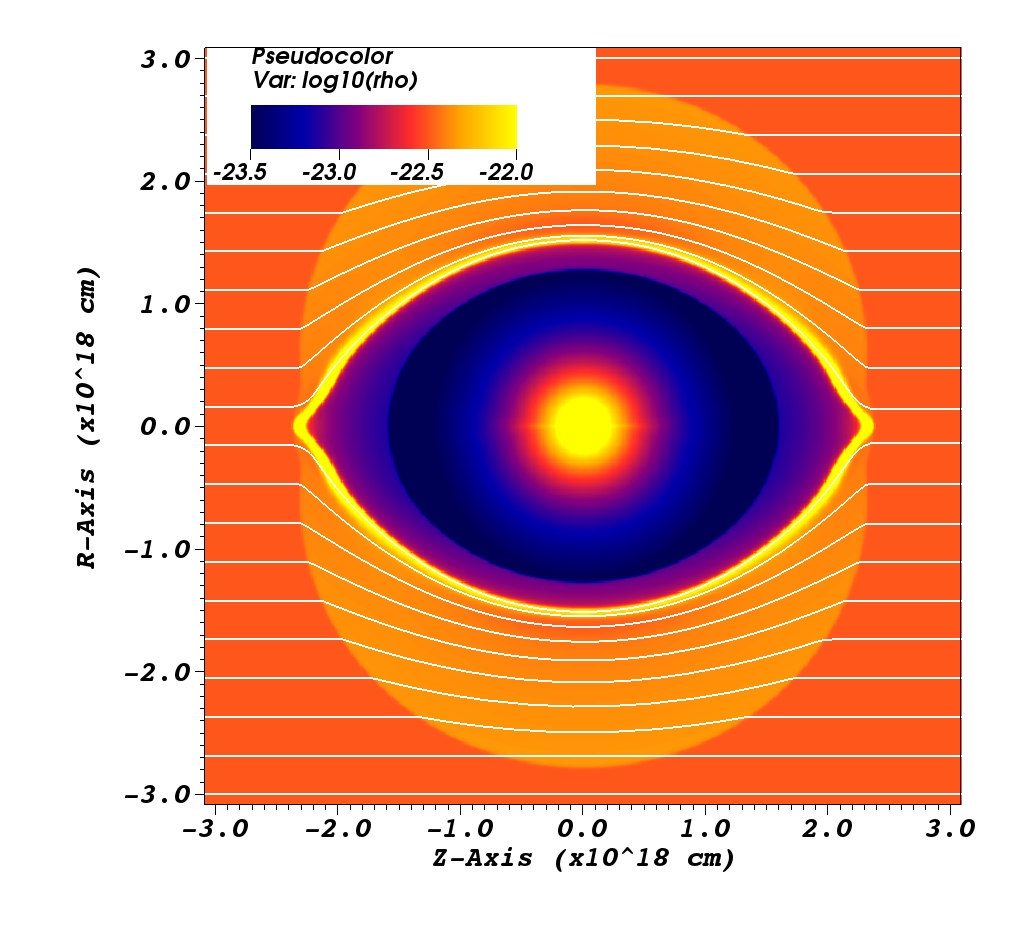}}
\subfigure
{\includegraphics[width=0.5\textwidth]{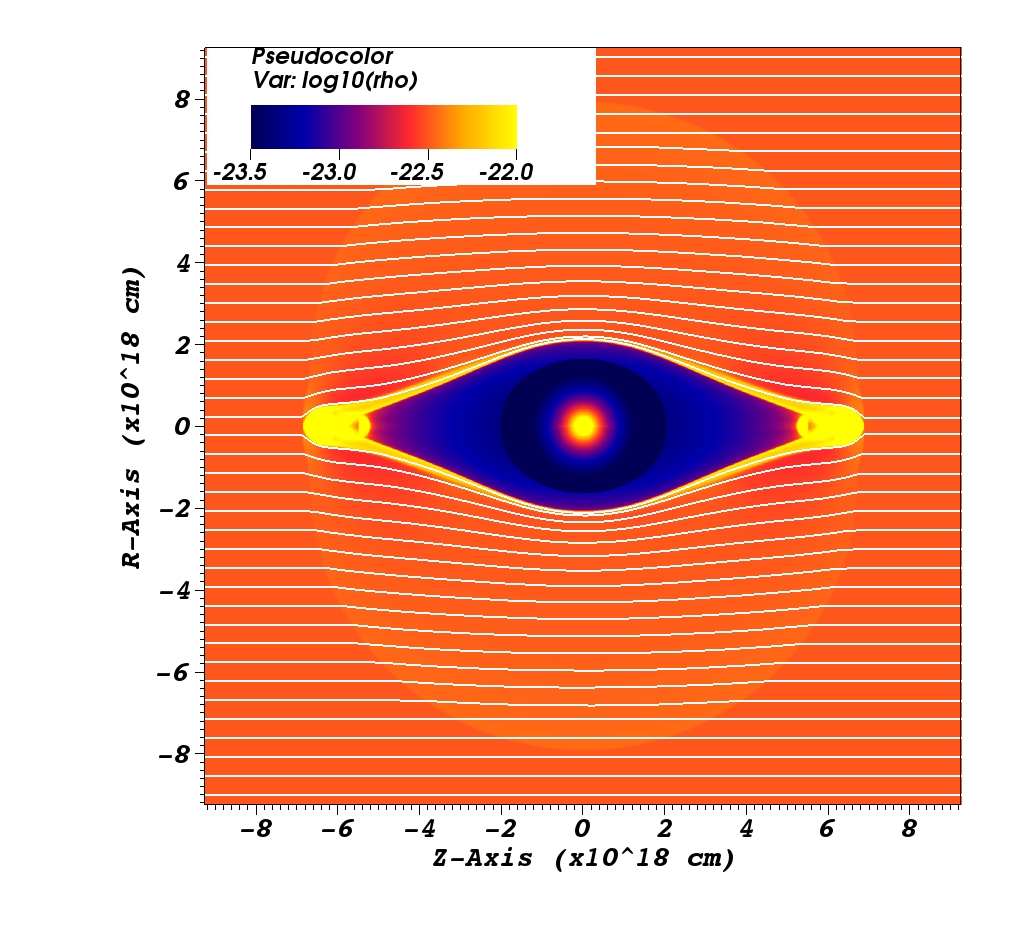}}}
}
\caption{Similar to Figs.~\ref{fig:B1}-\ref{fig:B3}, but for simulation~B4. 
The \emph{eye} shape starts after 190\,000 years (left) and the collimated flows appear after 730\,000~years (right).
Physical scales are identical to those in Figs.~\ref{fig:B1}-\ref{fig:B2}.}
 \label{fig:B4}
\end{figure*}

\section{Results}\label{sec-results}
\subsection{Wind expansion without a magnetic field}
With no magnetic field to constrain its expansion (see Fig.~\ref{fig:A1A2}), the wind blown bubble is spherical and qualitatively follows the evolution described analytically by 
\citet{Avedisova:1972} and \citet{Weaveretal:1977}. 
Starting from the star, the wind is initially free-streaming, with a constant radial velocity and a density that decreases with the distance squared. 
The wind then passes through the termination shock where it is thermalized and loses most of its kinetic energy to form a hot, low density bubble 
that expands outward due to its own thermal pressure. 
In doing so, it sweeps up a shell of interstellar matter, which move supersonically into the ISM. 
The shocked wind and the dense shell of shocked interstellar material are separated by a contact discontinuity. 
Quantitatively, the comparison with the analytical model is more difficult. 
The analytical models are based on the assumption of a purely adiabatic interaction (no radiative cooling included) between a fast 
($\sim$1000$\kms$) wind and the interstellar medium as can be expected for the main sequence phase of massive stars. 
Obviously this is not the case here as the AGB wind has a wind velocity that is two orders of magnitude smaller and a much higher density.  
As a result, our ``hot'', shocked wind bubble has a temperature of less than 10\,000\,K, rather than the $10^6-10^8$\,K expected for massive star bubbles. 
Nevertheless, Eq.~21 from \citet{Weaveretal:1977} puts the radius of the outer shell for both simulations at 0.5\,pc after 0.1\,Myr and  2\,pc after 1\,Myr, 
which matches the results shown in Fig.~\ref{fig:A1A2}. The fact that simulation~A2 consistently has a slightly smaller bubble than simulation~A1, 
even though analytically they should have the same radius, can be explained by energy loss through radiative cooling, 
which is more effective at the higher densities in this model. 

Initially both the forward and reverse shock are at least partially radiative, causing a high rate of compression, especially in the shell of swept-up ISM. 
In the case of simulation~A2, which has higher densities, the shell is thin enough to show linear thin-shell instabilities \citep{Vishniac:1983}, 
which are caused by the directional asymmetry between the ram pressure acting on the outer boundary of the shell and the thermal pressure exerted on the inner boundary of the shell by the shocked AGB wind. 
As the bubble expands the expansion velocity decreases, reducing the compression of the shell, which becomes thicker.  
Moreover, the density decreases, which reduces the radiative cooling rate \citep{Stevensetal:1992}. 
As a result, both the forward and reverse shock become purely adiabatic and the thin-shell instabilities disappear.

\subsection{The influence of the interstellar magnetic field}
Even with a weak (5\muG) magnetic field (Figs.~\ref{fig:B1}-\ref{fig:B4}), the expansion of the circumstellar bubble quickly becomes a-spherical 
because the field reduces the expansion in the direction perpendicular to the field. 
Even before the wind termination shock reaches the critical point ($\rb$), where ram pressure and magnetic pressure are in balance, 
the circumstellar shell is distorted into a shape that starts as ovoid (with the long axis parallel to the magnetic field), 
reminiscent of the results found for super bubbles by \citet{Tomisaka:1990,Tomisaka:1992} and \citet{Ferriere:1991}. 
The magnetic field pressure resists the compression, which, in the non-magnetic models, confines the shell of shocked ISM. 
As a result, rather than a highly compressed shell of interstellar matter, a relatively small distortion 
of the interstellar medium density moves outward in the direction perpendicular to the field lines. 
This is most visible in the left panels of Figs.~\ref{fig:B3} and \ref{fig:B4}. 
The interstellar medium is compressed, just ahead of the contact discontinuity, causing a sub-Alfv{\'e}nic wave to
move outward into the ISM, causing a distortion of the field lines. 
Behind the contact discontinuity, the shocked wind material, which is stopped from expanding piles up against the ISM. 
This is fundamentally different from the non-magnetic case, where the dense shell consisted exclusively of interstellar matter. 
In the magnetic model, the shocked wind itself contributes to the shell which might lead to a stronger infrared signal because of the high dust-content of the AGB wind. 
In reality, this distinction would be less clear than in the numerical model, 
because the dust grains might penetrate from the shocked wind into the shocked ISM because of their large inertia, relative to the gas \citep{vanMarleetaldust:2011}. 

Over time the asymmetry increases and forms tapering points along the long axis (the $z$-axis in our simulations). 
This shape, ovoid, with sharp points, eventually turns \emph{eye}-like as the points becomes elongated. 
This is caused by the magnetic tension force, which attempts to straighten the field lines by reducing local 
curvature. 
The moment this occurs varies from one model to another, depending on the input parameters 
as demonstrated in Table~\ref{tab:shape}, which shows the time at which the shape becomes 
\emph{eye}-like as well as the diameter along the long axis of the bubble at that moment. 
Although \emph{eye}-like is a subjective judgement, we have tried to define a consistent criterion: 
in order to be considered \emph{eye}-like, the shell has to have A) at least 1.5:1 diameter ratio between the long and short axes, and B) tapering points. 
The actual shape of the nebulae as defined in Table~\ref{tab:shape} 
for the onset of the \emph{eye}-like shape is demonstrated in the left panels of Figs.~\ref{fig:B1}-\ref{fig:B4}, 
which show the gas density and the shape of the magnetic field-lines. 

Eventually, the bubble expands to the point where the magnetic field pressure dominates over the ram pressure of the wind (the termination shock reaches $\rb$). 
Once this occurs, all expansion perpendicular to the field lines ceases. 
Expansion parallel to the field continues, because the stellar wind continues the feed mass and energy into the bubble. 
This leads to the formation of elongated structures consisting of collimated, outflows on either side of the circumstellar bubble, 
extending parallel to the magnetic field lines. 
Again, the age and largest diameter of the bubble at the start of the collimated flows is noted in Table~\ref{tab:shape}. 
The density of the circumstellar nebulae as well as the shape of the interstellar magnetic field lines is shown in the right-side panels of Figs.~\ref{fig:B1}-\ref{fig:B4}. 
Figure~\ref{fig:B1_end} shows the density and temperature of simulation~B1 after 1\,Myr, clearly demonstrating the behaviour of the collimated flows, which at this time 
have stretched to approximately 5\,pc on each side of the star.

The simulations with a 10\muG\, interstellar magnetic field (Figs.~\ref{fig:C1}-\ref{fig:C4}) follow the same pattern as for the 5\muG\, field. 
Because the magnetic field is stronger it starts to dominate the interaction in an earlier phase of the evolution, 
reducing the time and length-scales for the onset of the \emph{eye}-like shape (Table~\ref{tab:shape} and left panels of Figs.~\ref{fig:C1}-\ref{fig:C4}) 
as well as the formation of collimated flows (Table~\ref{tab:shape} and right panels of Figs.~\ref{fig:C1}-\ref{fig:C4}). 

When compared to the analytical prediction of the stand-off distance between the stellar wind and the magnetic field  
(Sect.~\ref{sec-analytic}) the \emph{eye}-like structures tend to be smaller, 
because they start to form in the time when the magnetic field influences the shape of the shell, 
but does not yet dominate completely over the ram pressure of the wind.
The scaling-dependence on the magnetic field strength seems to match the analytical approximation. 
As a pattern, size tends to depend more on mass loss rate than ISM density.
 
\begin{figure}
\FIG{
 \centering
\mbox{
\subfigure{\includegraphics[width=\columnwidth]{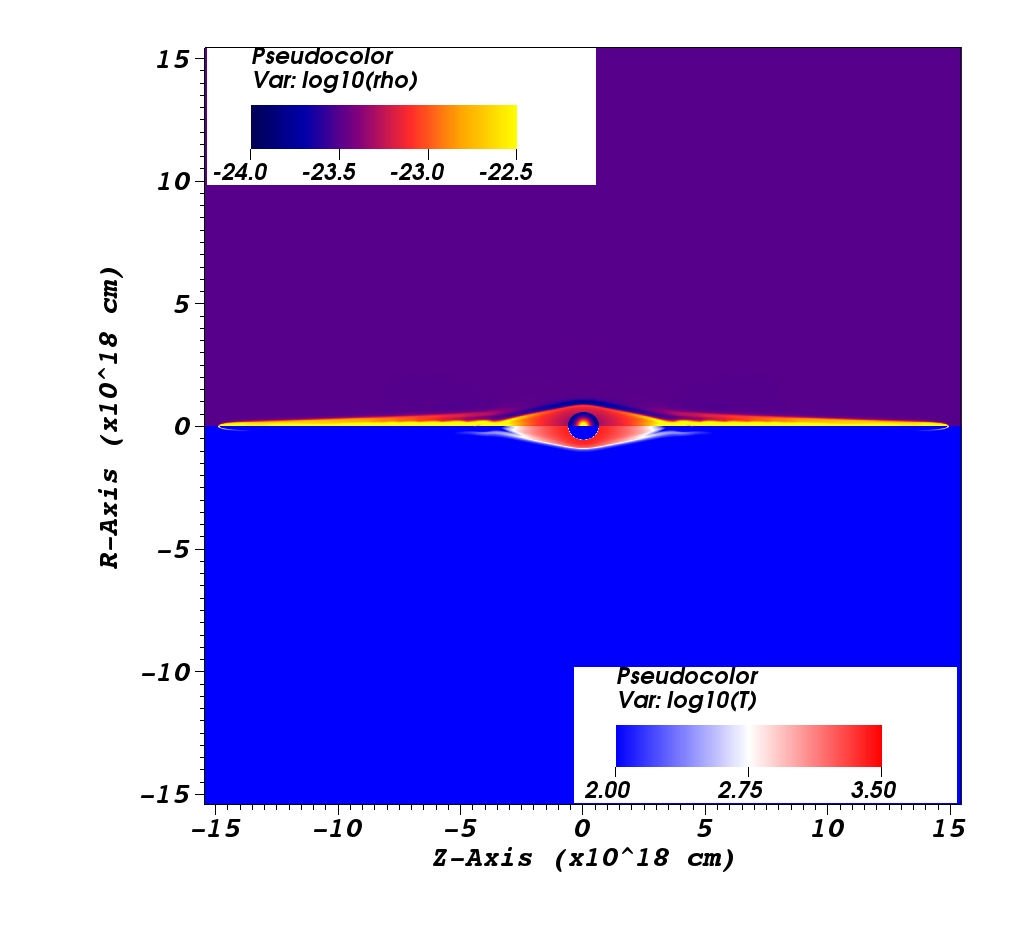}}}}
\caption{Density (top) and temperature (bottom) in [cgs.] units for simulation~B1 after 1\,Myr. The collimated flows have taken on 
a ``jet''-like appearance and stretch approximately 5\,pc in each direction.}
\label{fig:B1_end}
\end{figure}

\begin{figure*}
\FIG{
 \centering
\mbox{
\subfigure
{\includegraphics[width=0.5\textwidth]{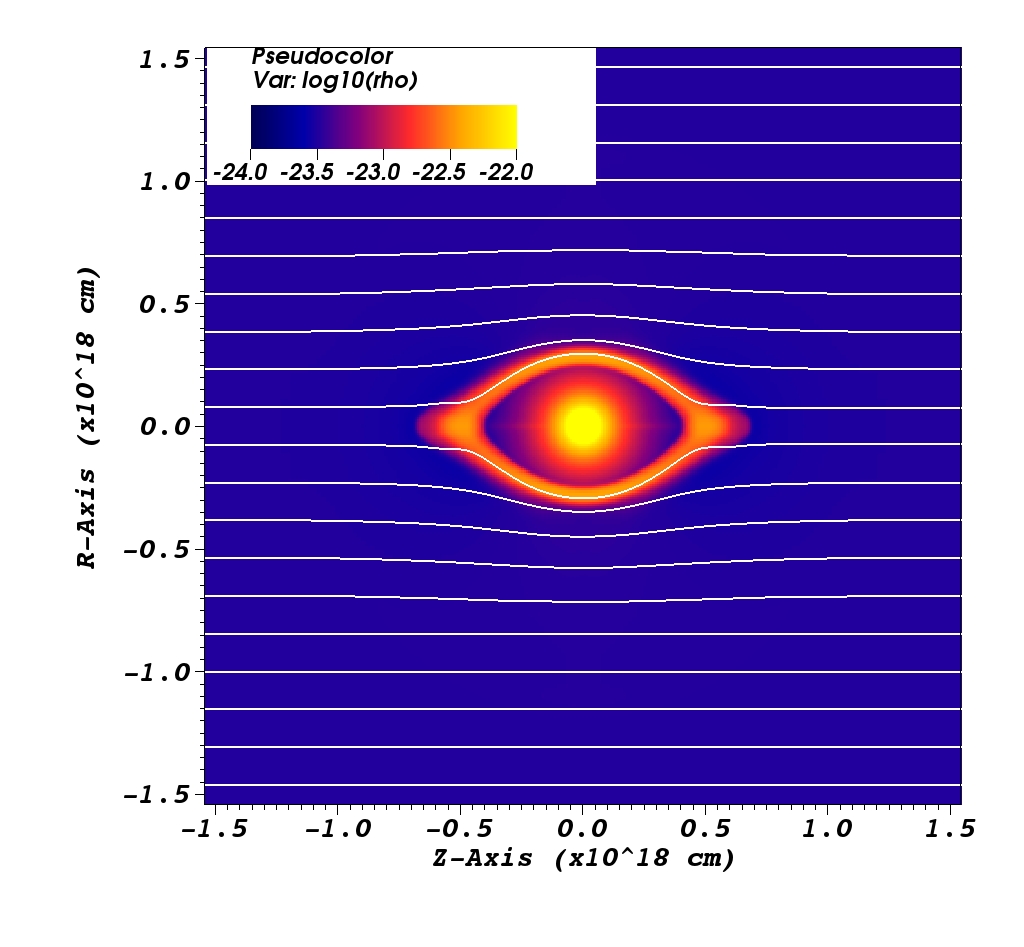}}
\subfigure
{\includegraphics[width=0.5\textwidth]{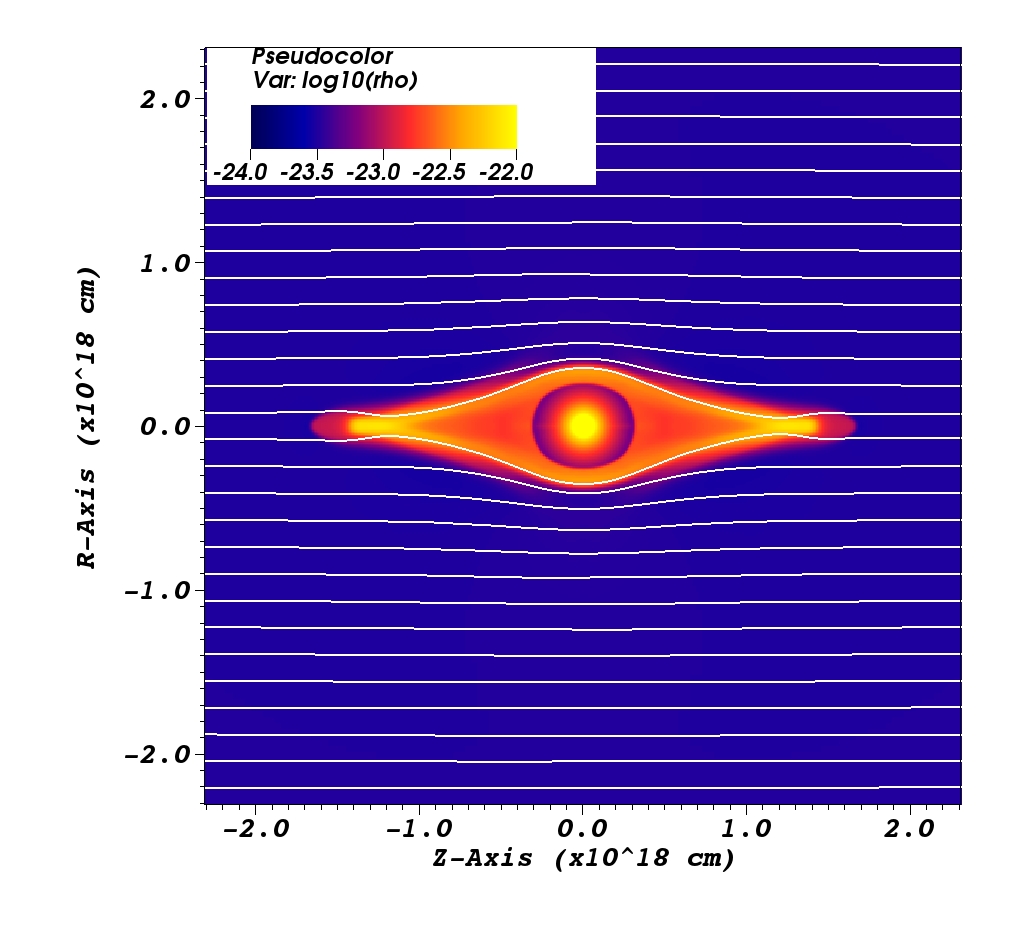}}}
}
\caption{Gas density and magnetic field lines for simulation~C1 after 20\,000 years (left, the onset of the \emph{eye} shape) 
and 60\,000 years (right, the beginning of the jet formation.)
Physical scales are 1$\times$1~pc (left panel) and 1.5$\times$1.5~pc (right panel).}
 \label{fig:C1}
\end{figure*}

\begin{figure*}
\FIG{
 \centering
\mbox{
\subfigure
{\includegraphics[width=0.5\textwidth]{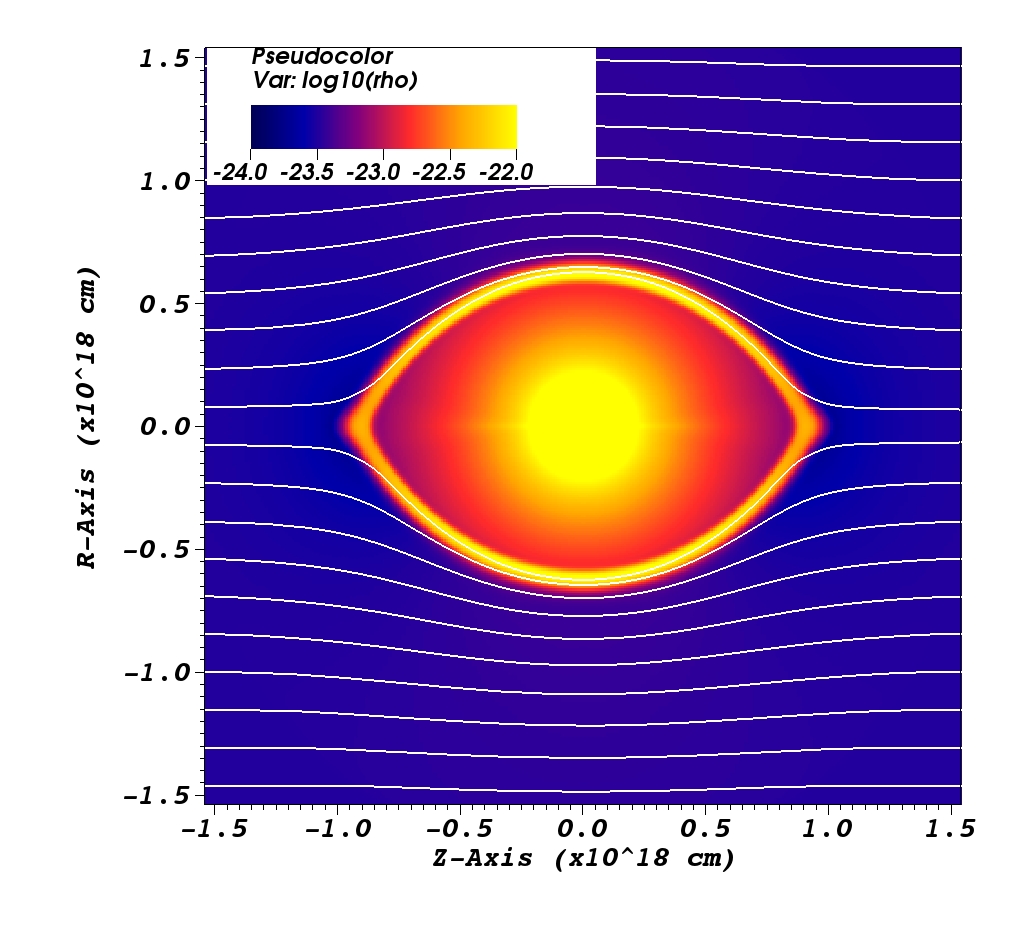}}
\subfigure
{\includegraphics[width=0.5\textwidth]{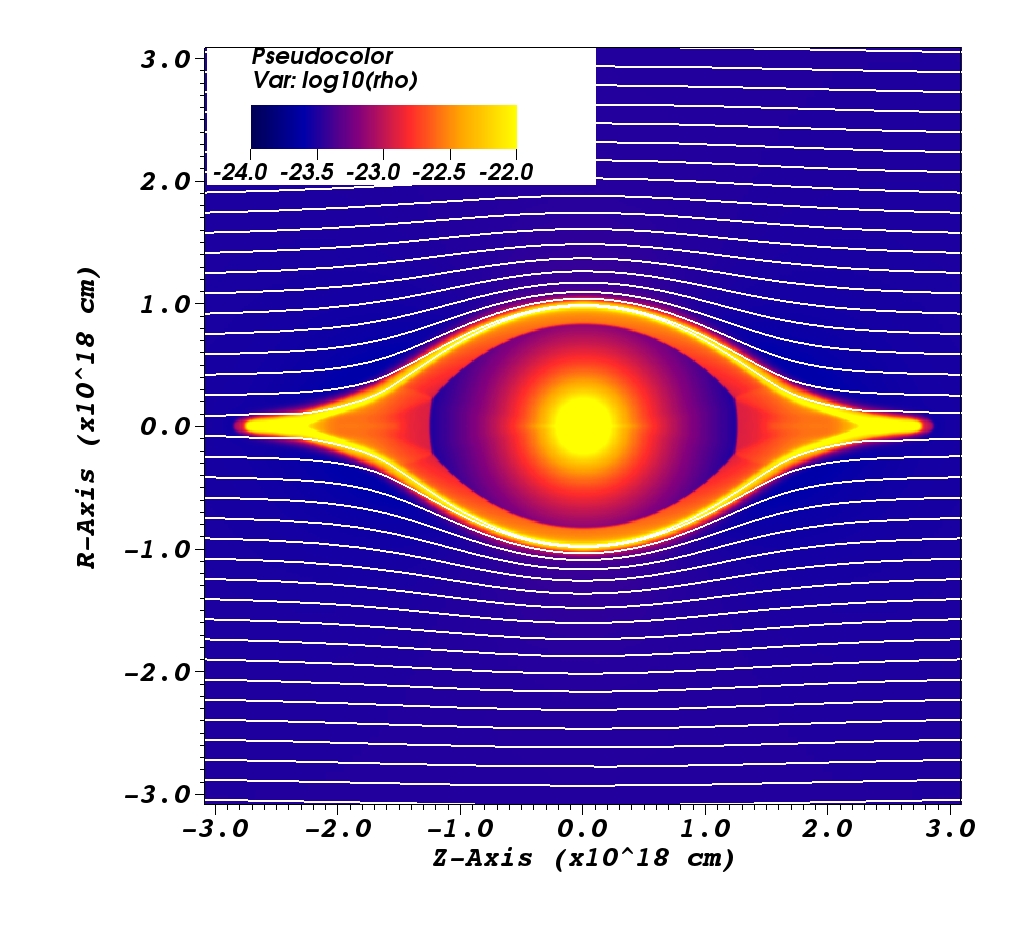}}}
}
\caption{Similar to Fig.~\ref{fig:C1}, but for simulation~C2. 
The \emph{eye} shape starts after 30\,000 years (left) and the collimated flows appear after 110\,000~years (right).
Physical scales are 1$\times$1~pc (left panel) and 2$\times$2~pc (right panel). 
Note the scaling difference between the right panel of this figure and the right panel of Fig.~\ref{fig:C1}.}
 \label{fig:C2}
\end{figure*}

\begin{figure*}
\FIG{
 \centering
\mbox{
\subfigure
{\includegraphics[width=0.5\textwidth]{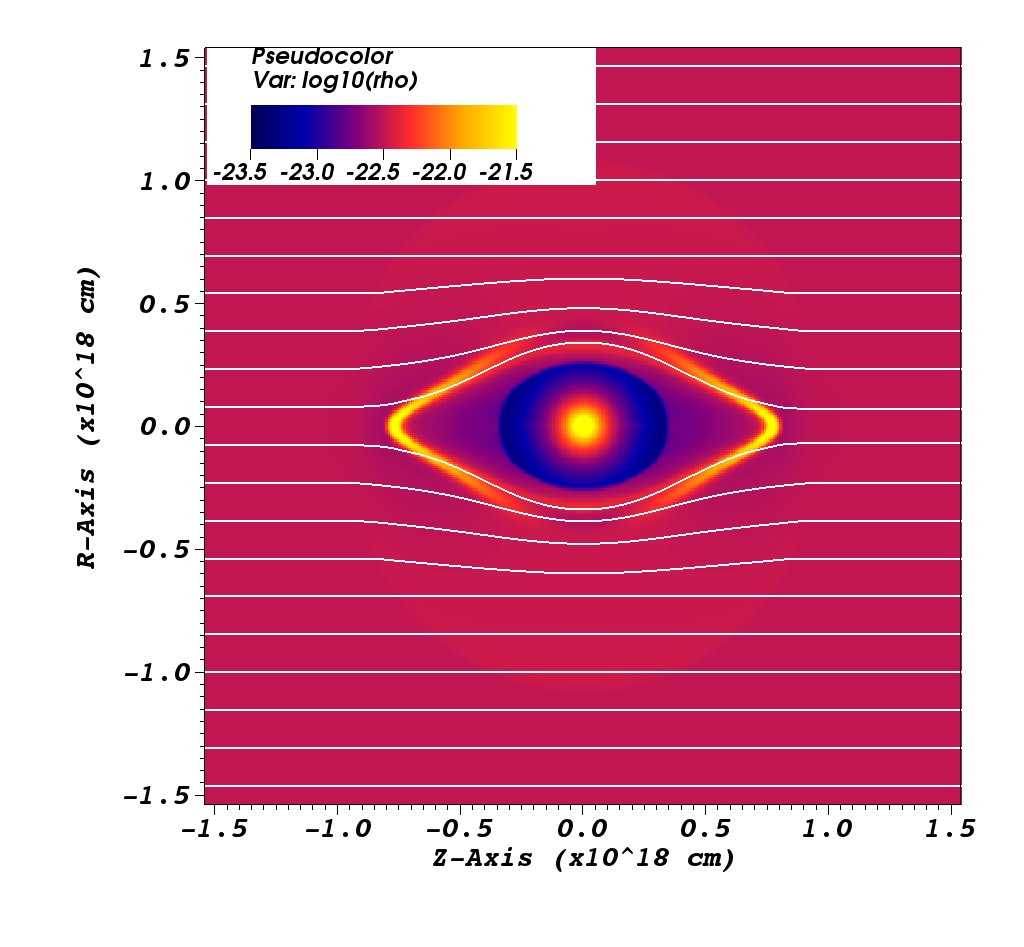}}
\subfigure
{\includegraphics[width=0.5\textwidth]{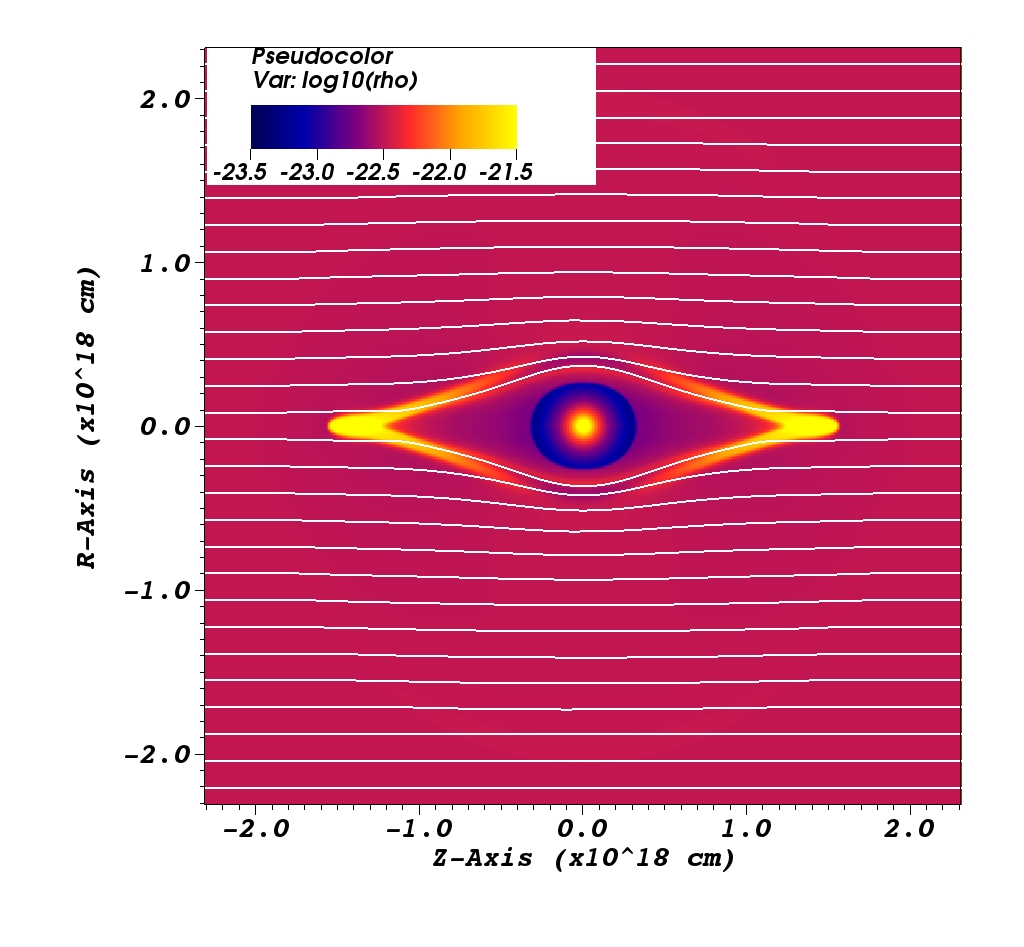}}}
}
\caption{Similar to Figs.~\ref{fig:C1}-\ref{fig:C2}, but for simulation~C3. 
The \emph{eye} shape starts after 50\,000 years (left) and the collimated flows appear after 110\,000~years (right).
Physical scales identical to those in Fig.~\ref{fig:C1}.}
 \label{fig:C3}
\end{figure*}

\begin{figure*}
\FIG{
 \centering
\mbox{
\subfigure
{\includegraphics[width=0.5\textwidth]{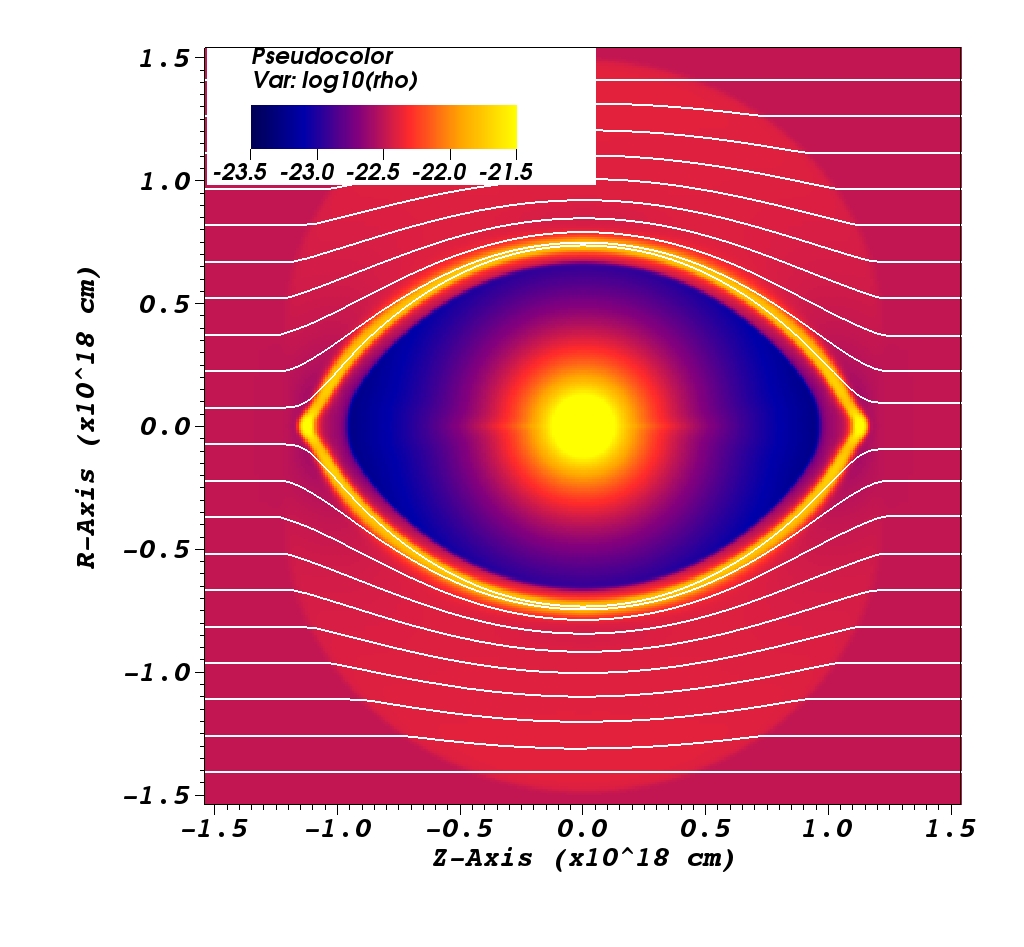}}
\subfigure
{\includegraphics[width=0.5\textwidth]{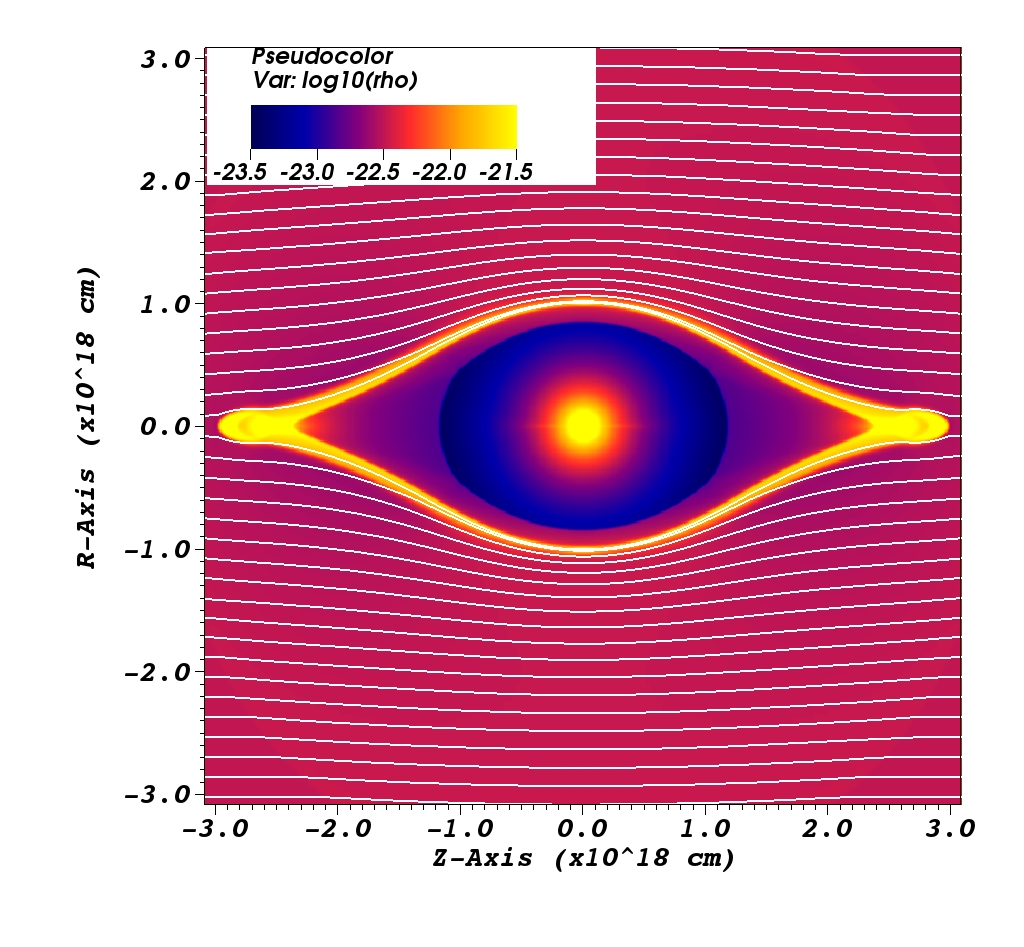}}}
}
\caption{Similar to Figs.~\ref{fig:C1}-\ref{fig:C3}, but for simulation~C4. 
The \emph{eye} shape starts after 60\,000 years (left) and the collimated flows appear after 190\,000~years (right).
Physical scales are identical to thos in Fig.~\ref{fig:C2}.}
 \label{fig:C4}
\end{figure*}

\subsection{3-D vs. 2-D}
\label{sec-3D}
To determine whether the use of a 2.5-D, cylindrically-symmetric grid influences the shape of the bubble, 
we run a single model in 3-D, using a carthesian grid and the same input parameters as for Simulation~B1.  
For this 3-D model we reduce the resolution by a factor 4 and limit the size of the grid in the directions perpendicular to the field to decrease the computation time. 
Fig.~\ref{fig:3D} shows cross-sections of the result after 40\,000 and 300\,000~years (the onset of the \emph{eye}-like shape and the start of the collimated flows, respectively.) 
Because of the lower resolution, the circumstellar shell is not as well resolved, but the result is qualitatively comparable to the 2.5-D model. 
The nebula initially takes the form of a ``lemon'', which would appear as an \emph{eye}-like shape (left and centre panels of Fig.~\ref{fig:3D}) when projected on the sky. 
During the later stages of the evolution the 3-D model forms the same collimated outflows as found in the 2.5-D simulations (right panel of Fig.~\ref{fig:3D}).

\subsection{Warm vs. cold ISM}
\label{sec-warmISM}
Figure~\ref{fig:E1} shows the result of simulation~E1 at the same moments in time as Fig.~\ref{fig:B1} to demonstrate the effect of a 
warm ISM (8000\,K). Both the \emph{eye}-shape (left panel) and the collimated flows (right panel) are clearly visible. 
However the nebula is significantly smaller. 
After 40\,000 years (left panel), the major axis of the nebula is only 0.45\,pc, rather than the 0.67\,pc found for the cold ISM (simulation~B1). 
Similarly, after 300\,000 years (right panel) the major axis is approximately 2\,pc, as opposed to 4\,pc for simulation~B1 
This reduction in size is caused by the thermal pressure of the ISM, which constrains the expansion.

\begin{figure*}
\FIG{
 \centering
\mbox{
\subfigure
{\includegraphics[width=0.33\textwidth]{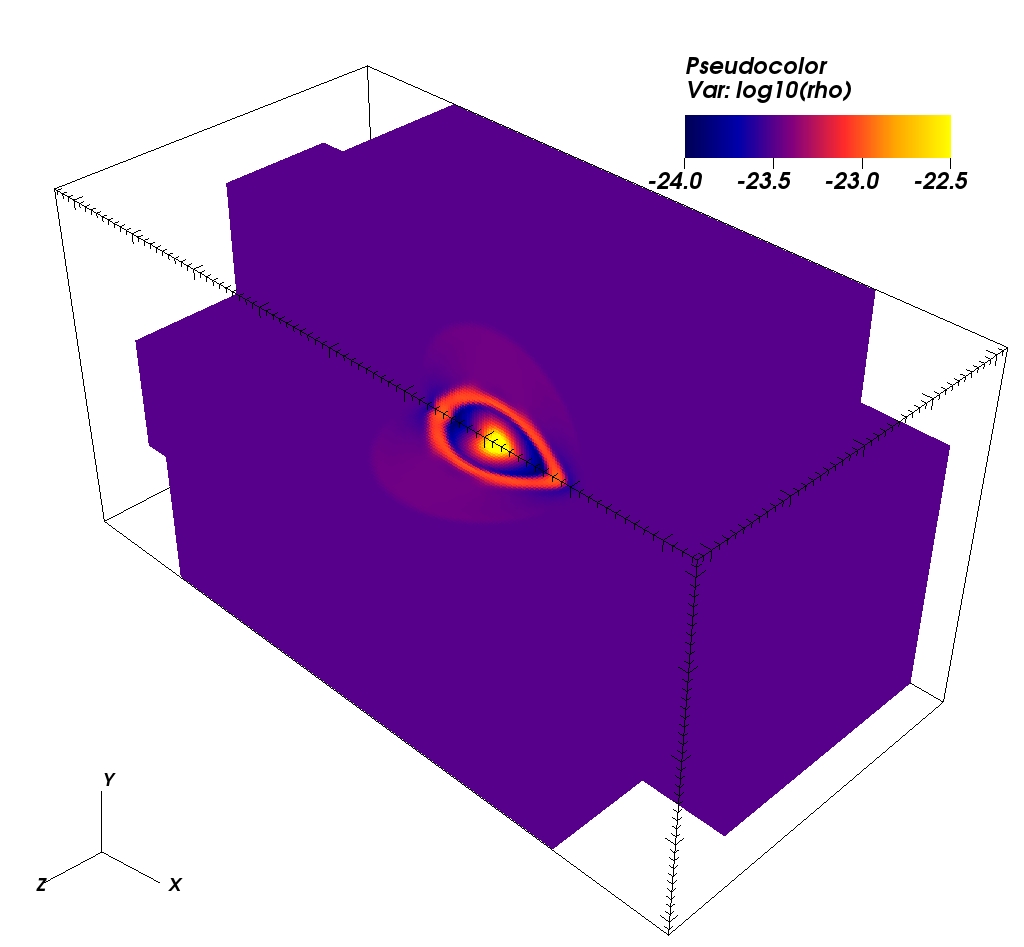}}
\subfigure
{\includegraphics[width=0.33\textwidth]{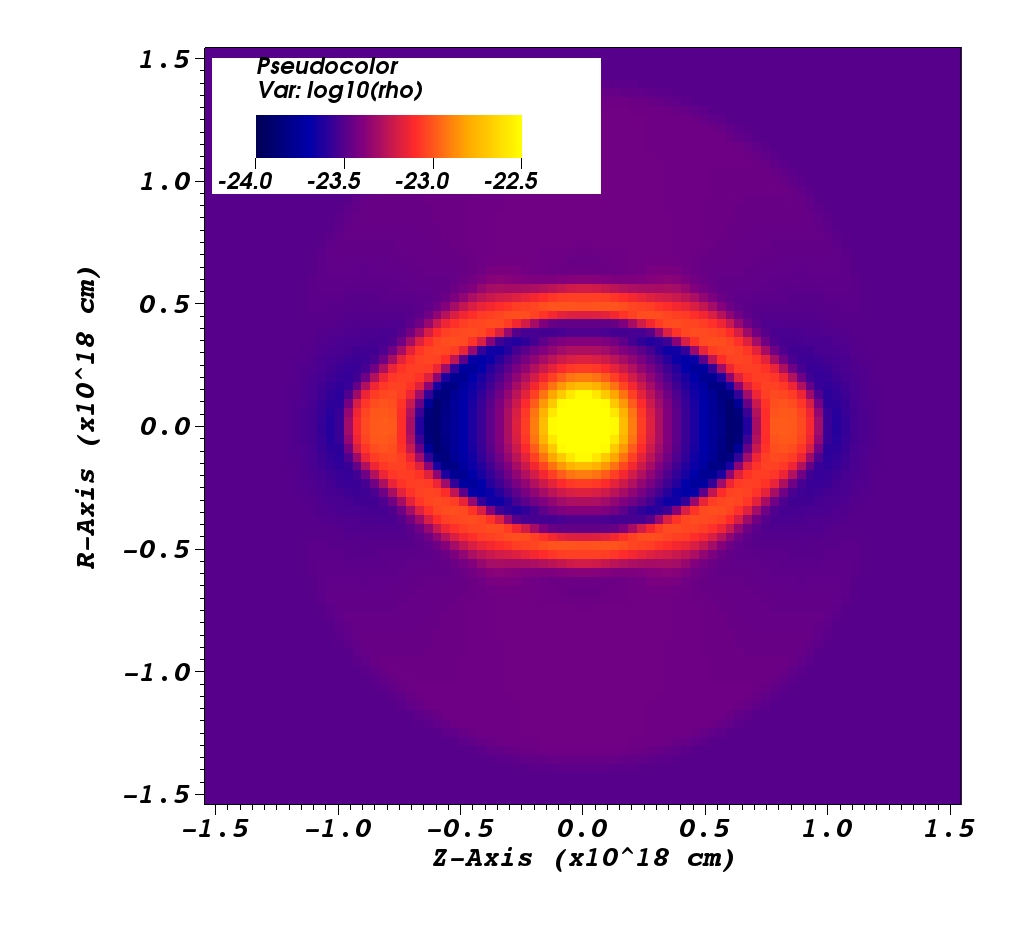}}
\subfigure
{\includegraphics[width=0.33\textwidth]{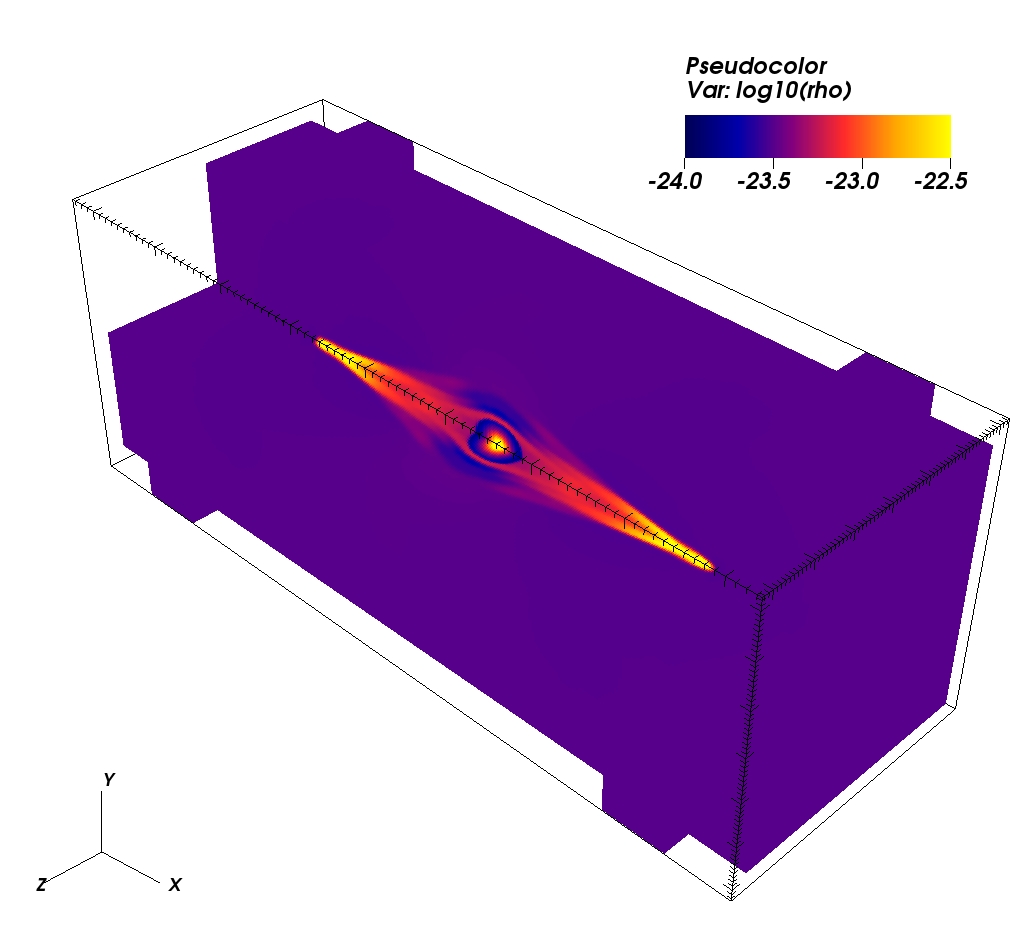}}}
}
\caption{3-D results for simulation~D1, showing the circumstellar density after 40\,000 (left and centre) and 300\,000~years (right), respectively. 
The centre panel shows a slice through the 3D data-cube that can be compared directly to the left panel of Fig.~\ref{fig:B1}. 
Although the shell is not as well resolved as for the 2.5-D models, both the \emph{eye}-like shape and the collimated flows are clearly visible. 
This confirms the validity of the 2.5-D models.}
 \label{fig:3D}
\end{figure*}

\begin{figure*}
\FIG{
 \centering
\mbox{
\subfigure
{\includegraphics[width=0.5\textwidth]{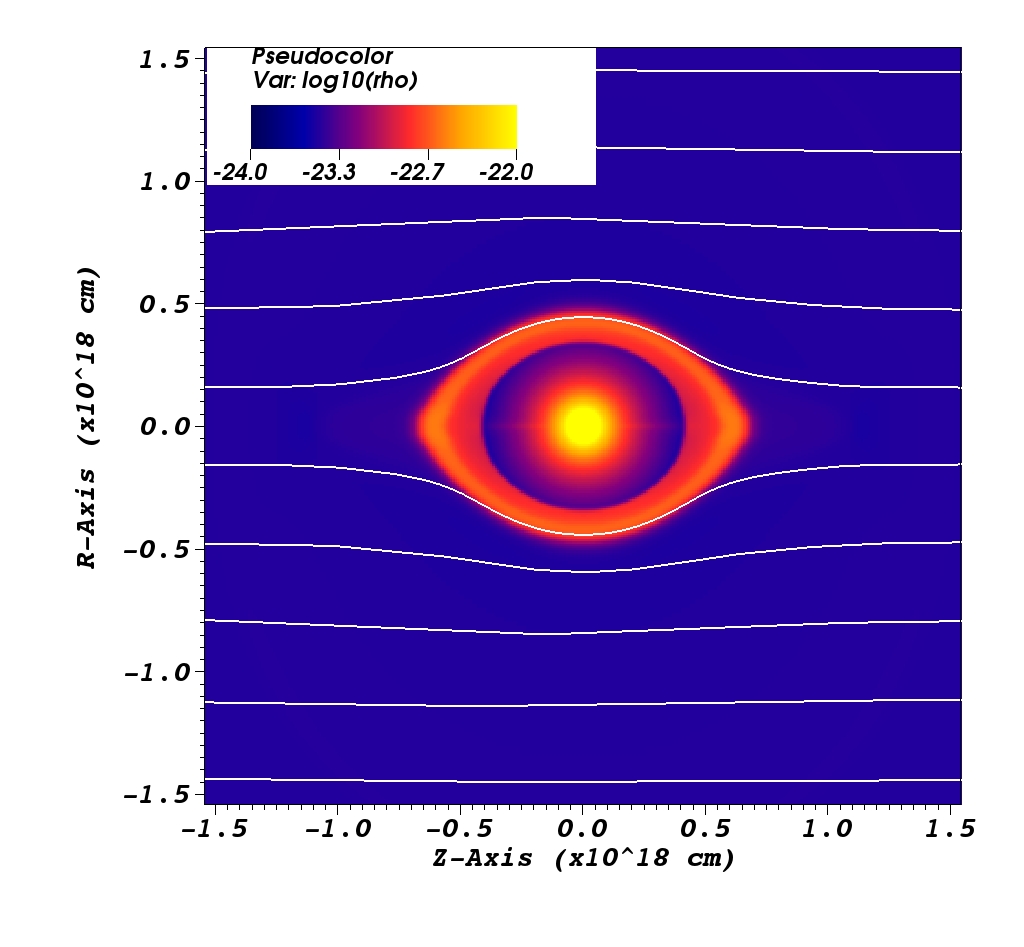}}
\subfigure
{\includegraphics[width=0.5\textwidth]{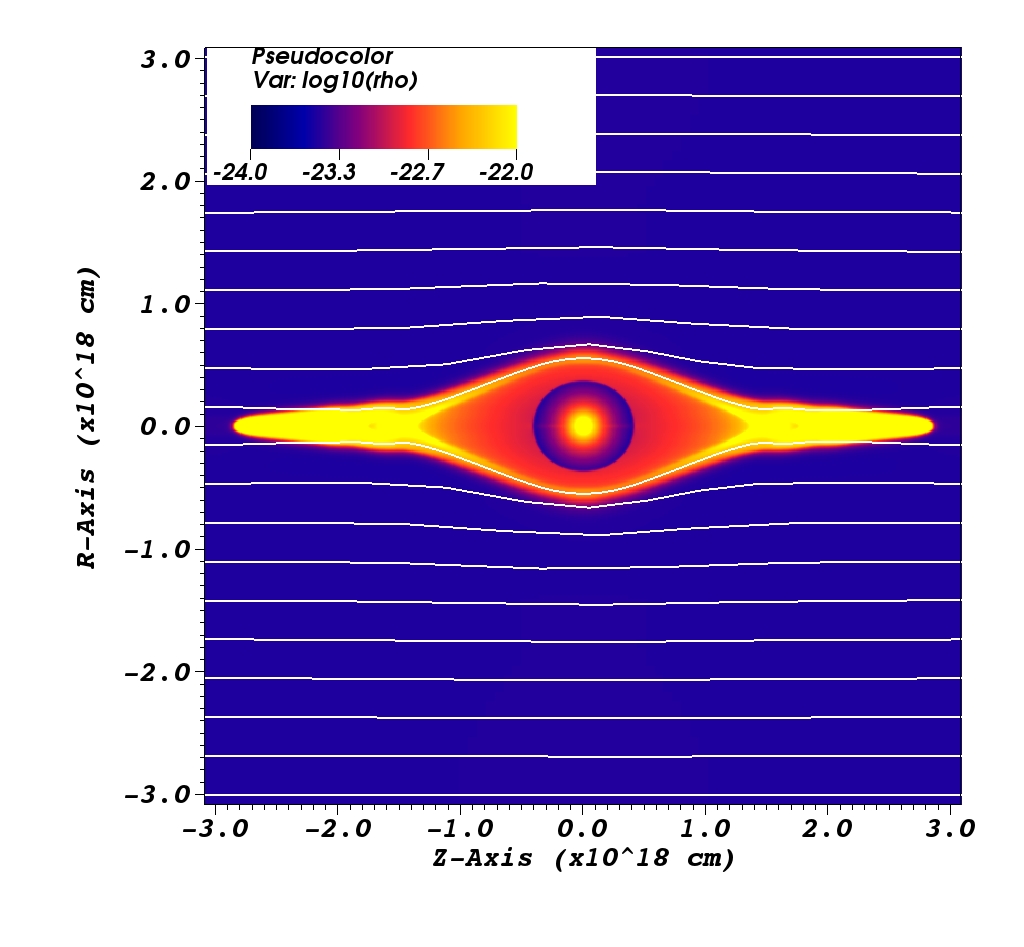}}}
}
\caption{Similar to Fig.~\ref{fig:B1} for simulation~E1 at the same moments in time. 
Physical scales are 1$\times$1~pc (left panel) and 2$\times$2~pc (right panel) and the colour bar has been adjusted to show the density structure. 
The physical size of the nebula is significantly smaller than for the cold ISM.
}
 \label{fig:E1}
\end{figure*}

\section{Discussion}
\label{sec-discussion}
\subsection{Shape}
The question is not ``How can we obtain an \emph{eye} shape?'', but, rather ``Why don't we see more of them?''
As long as you have an AGB wind interacting with an interstellar magnetic field, sooner or later the bubble will appear \emph{eye}-like. 
However, in order for us to observe it, it also has to be visible. 
Using \citet[Eq.~5.44 from][]{Tielens:2005}, we find that for a stellar luminosity of $10^3\lso$ and a grain size of 0.1\micron, 
carbon based dust grains will have a temperature of approximately 55\,K at a distance of 0.1\,pc, 
whereas at 1.0\,pc the dust temperature will be reduced to about 25\,K, which is approximately the same temperature as the dust in the ISM \citep{Tielens:2005}. 
As a result, the shell may not be visible against the background of the interstellar medium, though the difference between the carbon-rich dust of the AGB wind and 
the oxygen rich interstellar dust may still be observable. 
The \emph{eye}-shaped nebulae in our simulations tend to be larger than the three observed \emph{eyes} and may be unobservable for part of all of their \emph{eye}-like phase. 
Comparison between the cross-sections in Table~\ref{tab:shape} with the semi-major axes listed in Table~\ref{tab:targets} shows 
that at the onset of the \emph{eye}-like phase, the simulation results are already as large as, if not larger than, the observed nebulae. 
The fact that a warm ISM reduces the size of the nebula without compromising the \emph{eye}-shape, as demonstrated in Fig.~\ref{fig:E1},  
may indicate that \emph{eye}-like nebulae are more likely to be found in regions with a warm ISM.

A second limitation is caused by stellar motion. 
If the star moves supersonically through the ISM, this will eventually determine the shape of the circumstellar shell, 
which will form a \emph{fermata}-type bow shock \citep{Coxetal:2012}.  
An \emph{eye}-like shape can only occur as long as the wind is still expanding in all directions, before it reaches a ram pressure balance with the ISM. 
Assuming the parameters of Simulation~B1 and a stellar motion of 20$\kms$. the stand-off distance of the bow shock ($\rd$) would be 0.063\,pc \citep{Wilkin:1996}, 
an order of magnitude smaller than the initial \emph{eye}-like structure (Table~\ref{tab:shape}). 
Therefore, if the star as modelled in Simulation~B1 were to travel at 20$\kms$, it would never form an \emph{eye}-like nebula, but  would, 
almost from the start appear to have a \emph{fermata}-type nebula. 
As a result, we can expect \emph{eye}-like structures to occur primarily around stars in a low density ISM. 
This will delay the formation of a bow shock and allow the star to form a nebula that is initially \emph{eye}-like, before it succumbs to the \emph{fermata} shape.

Changes in $\mdot$ or $\vinf$ have little effect on whether the ram pressure of the ISM or magnetic pressure dominates, since both
$\rd$ and $\rb$ scale with the square-root of these quantities. 
On the other hand, n$_\mathrm{ISM}$ and the spatial velocity (v$_\star$) are important, because they
determines $\rd$ ($\rb$ being independent of these parameters). 
 An additional limiting factor is T$_\mathrm{ISM}$ because the star needs to move supersonically in order to give rise to a bow shock. 
 Using the characteristics of the interstellar medium as defined by \citet[][Chapter~1]{Tielens:2005}, typical stellar motion (10-100$\kms$) \citep{BinneyMerrifield:1998} 
 is supersonic in cold ($\simle100\,$K) or warm ($\simeq8\,000\,$K) ISM, 
 but would be subsonic in the hot intercloud medium. 
 
Finally, we have to take into account the angle at which we observe the shape. 
The shell will only appear \emph{eye}-like if our line of sight is approximately perpendicular to magnetic field (cf.~Sect.~\ref{sec-3D}).  
When observed along the direction of the magnetic field, the shell will appear \emph{ring}-like. 
Unfortunately, this cannot be confirmed for the observed \emph{eyes}-shapes due 
to lack of knowledge of the local magnetic field direction (and strength).
The angle of orientation of the \emph{eyes} with respect to the  Galactic plane only provides a rough indication (Fig.~\ref{fig:eyes}).

\subsubsection{\emph{Ring}-like nebulae}
Although the results of Simulations~A1 and A2 reproduce the \emph{ring}-like nebulae observed with \emph{Herschel} \citep{Coxetal:2012}, 
this should not be interpreted as proof that the presence of \emph{ring}-like nebulae indicates the absence of an interstellar magnetic field. 
Not only can both \emph{eye}-like and \emph{fermata} nebulae appear as rings, depending on the angle of observation, but 
a \emph{ring}-like nebula can also be formed through a spherically symmetric outburst caused by a thermal pulse \citep{Villaveretal:2002,Villaveretal:2003}. 
If such an outburst occurs, the \emph{ring}-shaped ejecta will initially expand into the free-streaming AGB wind. 
During this stage the ring has no interaction with the interstellar medium. 
This is probably the explanation for the presence of a \emph{ring}-like structure inside the \emph{fermata}-type shell observed around 
\object{R~Scl} \citep{Coxetal:2012}. 

Even when interacting directly with the interstellar magnetic field, the outer shell may appear \emph{ring}-like as 
long as the ram pressure of the wind at the termination shock dominates over both the interstellar magnetic pressure and the magnetic tension force. 
This will occur if either the interstellar magnetic field is weak and/or the shell is still young, with the termination shock close to the star. 

\subsection{Compression}
Although the non-magnetic simulations (A1 and A2) start out with partially radiative shocks, over time the shocks become adiabatic 
and the compression is reduced. 
All simulations that include magnetic fields show a completely different development: 
the additional pressure exerted on the shell by the interstellar magnetic field causes the shocked wind to pile up at the contact discontinuity. 
The local density increases, which makes the radiative cooling more efficient. 
As a result, the reverse shock tends to remain at least partially radiative throughout the evolution of the circumstellar bubble.

\subsection{Gas-dust connection}
The observations done with \emph{Herschel} (Fig.~\ref{fig:eyes}) show the dust, rather than the gas. 
Numerical simulations of the \textalpha-Orionis bowshock \citep{vanMarleetaldust:2011,Coxetal:2012,Decinetal:2012} 
show that small dust grains are strongly coupled to the gas and will remain within the shocked gas region,  
most likely piling up at the contact discontinuity \citep{vanMarleetaldust:2011}.
Therefore, we can expect that, as long as the AGB wind, which contains the most dust forms an \emph{eye}-like shape, 
such a shape would also be visible in the infrared. 
N.B. These numerical models did not include any interaction between dust grains and a magnetic field. 
If the dust grains have an electric charge, caused moving through a plasma as well as the interaction with the interstellar UV field, 
they will be tied to the magnetic field, which will stop them from moving deep into the ISM. 
In general, the hydrodynamical interaction in the wind-ISM interaction region are not very energetic, 
due to the low velocity of the AGB wind. As a result, dust-grains are likely to survive the transition through the shock. 

\subsection{Would the outflows be visible?}
Over time, the collimated outflows extend along the field lines (Fig.~\ref{fig:B1_end}). 
However, no such structures have been observed, except, possibly, for the case of \VYUma\, (See Fig.~\ref{fig:eyes})  
and even there they have nowhere near the length of the collimated flows found in the 
simulations. 
There are several contributing factors that reduce the likelihood of observing these collimated flows: 
\begin{enumerate}
 \item They only occur in the later stages of the nebula evolution. 
 By this time the nebula is often distorted by the stellar motion, which will destroy the collimated flows, 
 just as it destroys the \emph{eye}-shape.
 \item The dust in the jets is too cold. The collimated flows only form in the later stages of the evolution of the bubble.
By that time, the bubble has expanded, increasing the distance to the central star, which reduces the dust temperature. 
\item The flows are kept collimated by the magnetic field tension and pressure. 
Owing to turbulence, the interstellar magnetic field is unlikely to be completely constant in strength and direction over long distances \citep{MinterSpangler:1996,Hanetal:2004} 
and any variation in the direction of the field would break up the collimated flows. 
\end{enumerate}

\begin{table}
      \caption[]{Age and size of eye-shapes.}
         \label{tab:shape}
         \begin{tabular}{lcccc}
            \hline \hline
            \noalign{\smallskip}
            Simulation      &  \multicolumn{2}{c}{Start \emph{eye}-shape\tablefootmark{a}}          &  \multicolumn{2}{c}{Start collimation\tablefootmark{a}}     \\
                            &  [$\times\,1000$\,yr] & [pc] &    [$\times\,1000$\,yr]  & [pc]        \\
            \noalign{\smallskip}
            \hline
            \noalign{\smallskip}              
            B1               &   40     & 0.67    &  300  &    4.0     \\
            B2               &   100    & 1.6     &  380  &    5.3     \\
            B3               &   120    & 0.83    &  430  &    2.3     \\
            B4               &   190    & 1.6     &  730  &    4.7     \\            
            \noalign{\smallskip}
            \hline
            \noalign{\smallskip}            
            C1               &   20     & 0.4     &  60   &   1.0       \\
            C2               &   30     & 0.65    &  110  &   2.0       \\
            C3               &   50     & 0.53    &  110  &   1.0       \\
            C4               &   60     & 0.8     &  190  &   2.0       \\                                
             \noalign{\smallskip}                       
            \hline
         \end{tabular}
      \tablefoot{   
 \tablefoottext{a}{The criteria used to define the \emph{eye}-shape and collimated flows are described in Sect.~\ref{sec-results}.}}      
   \end{table}

\section{Influence of interstellar magnetic fields on the evolution of Planetary Nebulae}
\label{sec-PN}
\subsection{Premise}
PNe are the product of the interaction between a fast, low density post-AGB wind and its AGB wind predecessor \citep{Kwoketal:1978}. 
This means that, the interstellar magnetic field, which shapes the AGB-wind bubble, can indirectly influence the shape of the PN. 
However, under most circumstances this is unlikely. 
The lifetime of a PN is limited by the evolution of its progenitor star, which will typically move from the initial post-AGB phase, 
to the white dwarf stage 
within approximately 10\,000\,years \citep[][and references therein]{Kwok:2000}. 
Once this happens, the PN, no longer driven by a stellar wind, or ionized by a strong radiation source becomes invisible. 
As our simulations A1 and A2 show, the typical AGB wind bubbles (disregarding the interstellar magnetic field), 
will reach radii of about 2\,pc within the lifetime of an AGB star. 
For PNe to interact with the interstellar medium at all, they would have to reach this distance within the 10\,000 years of its own lifetime, 
which requires a minimum velocity of approximately 200$\kms$. 
This is an order of magnitude higher than the typical expansion velocities found for PNe \citep{Terzian:1997}. 

One exception to this rule is the nearby PN \object{Sh-216}, which has an estimated radius of approximately 1.5\,pc. \citep{Tweedyetal:1995}. 
The expansion speed of the shell of this particular PN is estimated at 4$\kms$, less than a typical AGB wind velocity; 
whereas the shell should actually exceed the speed of the AGB wind that it is sweeping up. 
This  indicates that the shell is interacting directly with the ambient interstellar medium. 
\citet{Ransometal:2008} showed polarization data, probing the magnetic field in \object{Sh-216}, which indicate that the 
interstellar magnetic field is being pushed outward by the expanding shell. 
Similarly, \citet{SokerZucker:1997} showed that an interaction may have taken place between the outer halo of \object{NGC~6894} 
and the interstellar magnetic field. 

However, as shown in our simulations, the interstellar magnetic field can reduce the size of the AGB-wind bubble and, 
if the AGB wind bubble is small enough, the PN will start to interact directly with the ISM, once it has swept-up the AGB wind. 
For example, for simulation C1, the analytical value for $\rb$  is only 0.11\,pc, which is born out by the simulation result (See Fig.~\ref{fig:C1}). 
For a PN expansion speed of $20\kms$, this would mean that the PN reaches the wind termination shock after only 5500\,years, 
well within the lifetime of a typical PN. 
\citet{FalcetaGoncalvesMonteiro:2014} showed simulations of PNe interacting with interstellar magnetic fields and concluded 
that only very strong fields ($\simeq\,500$\muG) could significantly influence the shape of PNe. 
However, the study by \citet{FalcetaGoncalvesMonteiro:2014} focussed on bipolar PNe. 
In this work, we simply intend to demonstrate if, and how, the distortion of the AGB-wind bubbles by the interstellar magnetic field 
can influence the evolution of nebulae created in subsequent evolutionary phases. 

As an example, we choose to use simulation~C2, (high mass loss rate, low interstellar density and a 10\muG\, interstellar magnetic field) 
and use it as basis for a PN simulation. 
Using the result of this simulation at 0.1\,Myr as basis we launch a fast, post-AGB wind inside the wind blown bubble 
and follow its evolution over time. 
This is \emph{NOT} to suggest that we expect that the AGB phase would be as short as 0.1\,Myr. 
Our sole purpose is to demonstrate what would happen 
if a PN were to form in an \emph{eye}-shaped AGB-wind bubble. 
We estimate the post-AGB wind to have a mass loss rate of $10^{-7}\msoy$ and a terminal velocity of 1500$\kms$, reminiscent of the 
input parameters used by \citet{Garcia-Seguraetal:1999}. 
We do not consider the AGB-superwind \citep[][and reference therein]{Tanabeetal:1997,Kwok:2000}. 
Notwithstanding the effects of these additional factors are not trivial, they go beyond the scope of this paper. 

\subsection{Results} 
Figures \ref{fig:PNE1} and \ref{fig:PNE2} show the evolution of the PN at intervals of 2\,000 years. 
Initially, the PN is spherically symmetric (left panel of Fig.~\ref{fig:PNE1}.) 
This is to be expected, because the interaction is between the spherically-symmetric post-AGB wind 
and the spherically-symmetric, free-streaming AGB wind. 
because the swept-up shell is thin, highly compressed with thermal pressure from the shocked post-AGB wind on one side and ram-pressure from its own motion on the other, 
it is subject to linear thin-shell instabilities \citep{Vishniac:1983}. 
In addition, it shows sign of Rayleigh-Taylor instabilities, with clumps of high density material moving from the shell into the low-density gas of the shocked post-AGB wind. 
If observed at this particular moment in time, the outer, \emph{eye}-like shell may be visible as a halo around the PN. 
Such halos have been observed for several PNe, such as \object{Abell~30} \citep{Guerreroetal:2012} and \object{NGC~6751} \citep{Clarketal:2010}, 
but are usually spherical in shape and can also be attributed to mass ejecta from a thermal pulse, rather than the wind-ISM interaction.  
When the PN hits the termination shock of the the AGB wind bubble (right panel of Fig.~\ref{fig:PNE1}), it slows down. 
This is partially due to the fact that it has to overcome the thermal pressure of the shocked AGB wind, but primarily, 
because it now expands into a constant density medium, rather than the $1/r^2$ density profile of a free-streaming wind. 
As a result, the expansion velocity decreases over time, rather than remaining constant. 
Owing to the fact that the PN hits the termination shock first along the direction perpendicular to the magnetic field, 
this causes the PN to depart from spherical symmetry and become ovoid (left panel of Fig.~\ref{fig:PNE2}), 
a tendency that continues as it fills the AGB wind bubble,  giving the PN a somewhat \emph{eye}-like appearance, 
such as observed for \object{Abell~70} \citep{Miszalskietal:2012}. 

When the PN hits the outer shell of the AGB wind bubble, it starts to interact with the interstellar magnetic field. 
Initially, this means little for the expansion speed, which is driven by a relatively powerful wind 
($\rb$ for the post-AGB wind in a 10\muG\, magnetic field is approximately~1\,pc). 
However, the nature of the instabilities in the shell changes. 
The Rayleigh-Taylor instabilities on the inner side of the shell grow (right panel of Fig.~\ref{fig:PNE2}.) 
On the outer edge of the shell, the PN continues to display thin-shell instabilities.  
The wavelength of both thin-shell and Rayleigh-Taylor instabilities varies smoothly along the shell, 
with the  longest wavelength occurring where the expansion of the shell is perpendicular to the field lines. 
This phenomenon, which combines the influence of both field strength and direction on the wavelength of a perturbation 
was described analytically by, for example, \citet[][]{Junetal:1995,Breitschwerdtetal:2000}. 

Eventually, the PN encounters the tips of the \emph{eye}, while it expands along the $z$-axis (Fig.~\ref{fig:PNE3}). 
This changes the morphology of the PN, because sweeping up these high-density structures slows down the expansion. 
As a result, the PN takes on a new shape, somewhat reminiscent of a bi-polar nebula, as was 
also demonstrated, albeit for much stronger fields, by \citet{FalcetaGoncalvesMonteiro:2014}. 
However, one should keep in mind, that the symmetry axis here is the $z$-axis. 
Therefore, the nebula would only appear bi-polar when observed from the side. 
In 3-D the structure would actually be an ovoid that has been indented at the tips. 
Furthermore, The left and right panels of Fig.~\ref{fig:PNE3} show the PN after 15\,000 and 30\,000\,years respectively. 
By this time the central star would almost certainly have become a white dwarf, rendering the PN invisible. 
The outer shell is no longer thin, because the magnetic pressure from the compressed field 
counteracts the compression. 
This results in a disappearance of the instabilities. 
The general shape of the nebula strongly resembles the results of \citet{Tomisaka:1990,Tomisaka:1992} and \citet{Ferriere:1991} 
for the expansion of superbubbles into an interstellar magnetic field.

\begin{figure*}
\FIG{
 \centering
\mbox{
\subfigure
{\includegraphics[width=0.5\textwidth]{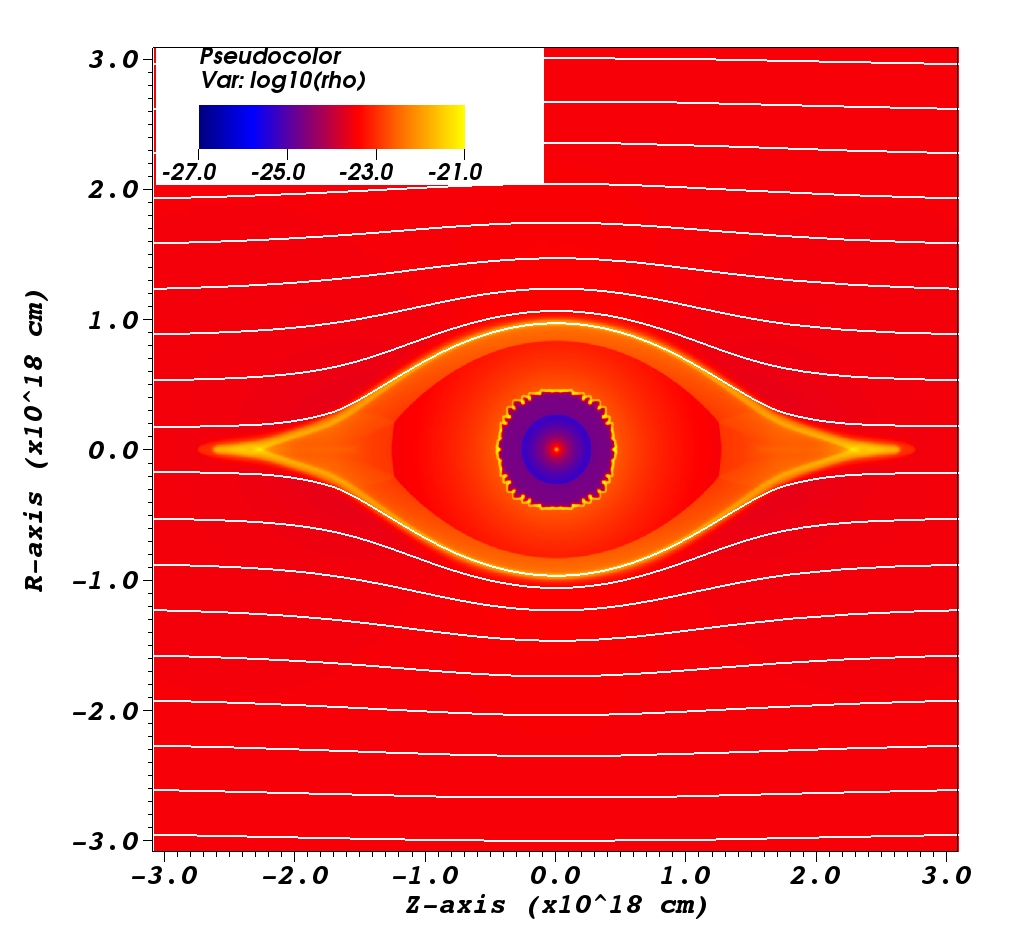}}
\subfigure
{\includegraphics[width=0.5\textwidth]{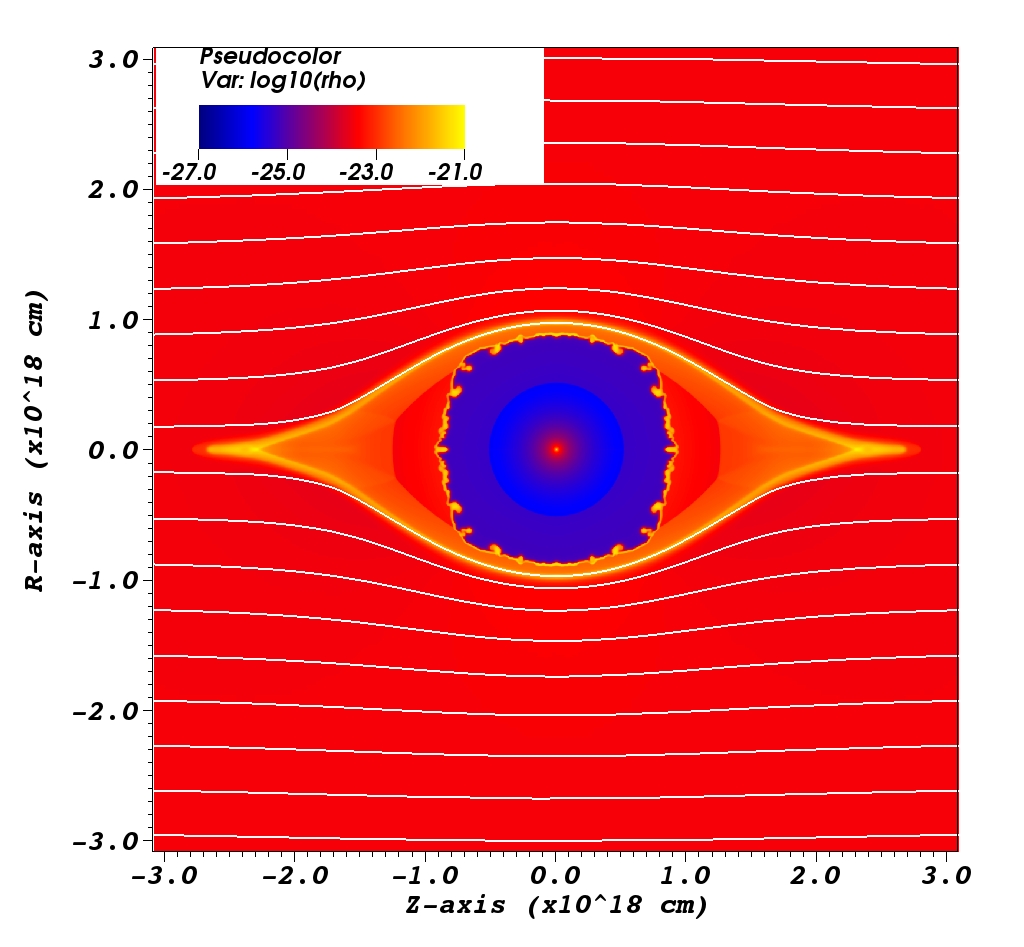}}}
}
\caption{Evolution of a PN in an eye-shaped AGB-wind bubble, showing the gas density [cgs.] 
and magnetic field lines at 2\,000 (left) and 4\,000 (right) years after the onset of the post-AGB phase. 
The PN, which starts out spherically (left) starts to feel the influence of the eye-like shape 
when it encounters the termination shock of the AGB wind (right).}
 \label{fig:PNE1}
\end{figure*}

\begin{figure*}
\FIG{
 \centering
\mbox{
\subfigure
{\includegraphics[width=0.5\textwidth]{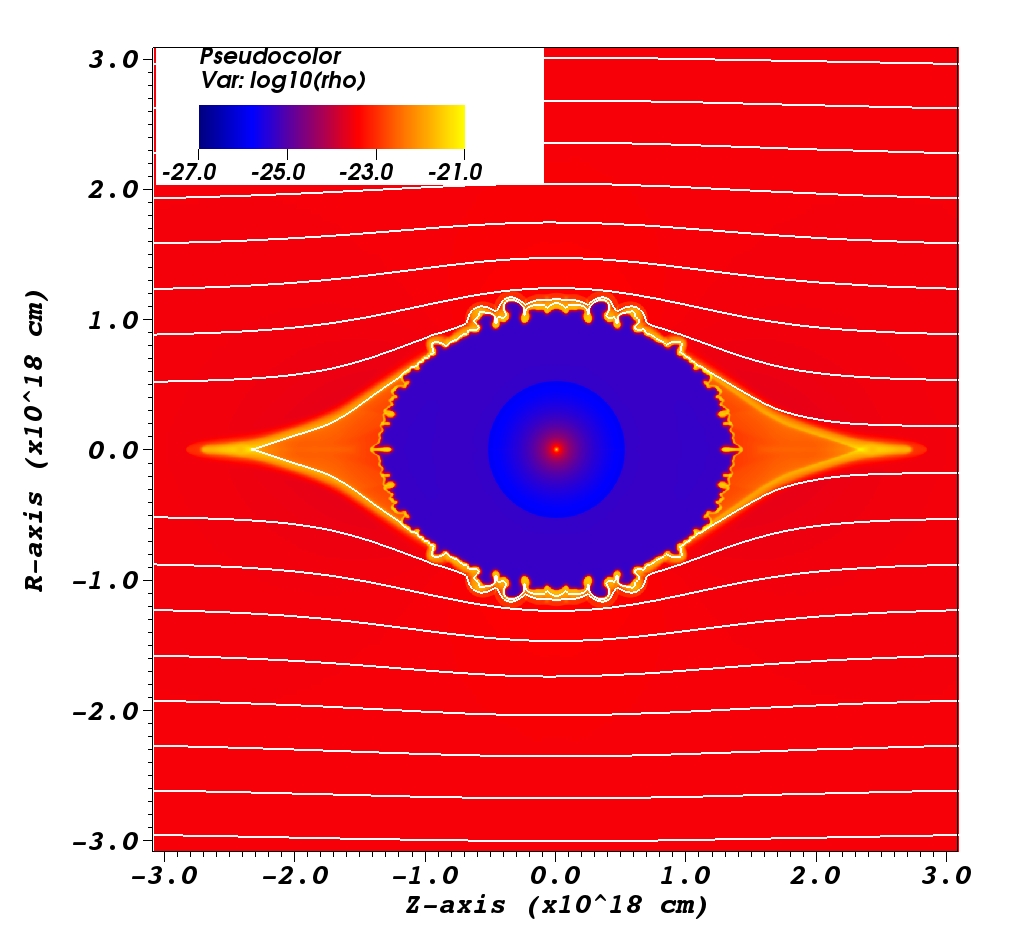}}
\subfigure
{\includegraphics[width=0.5\textwidth]{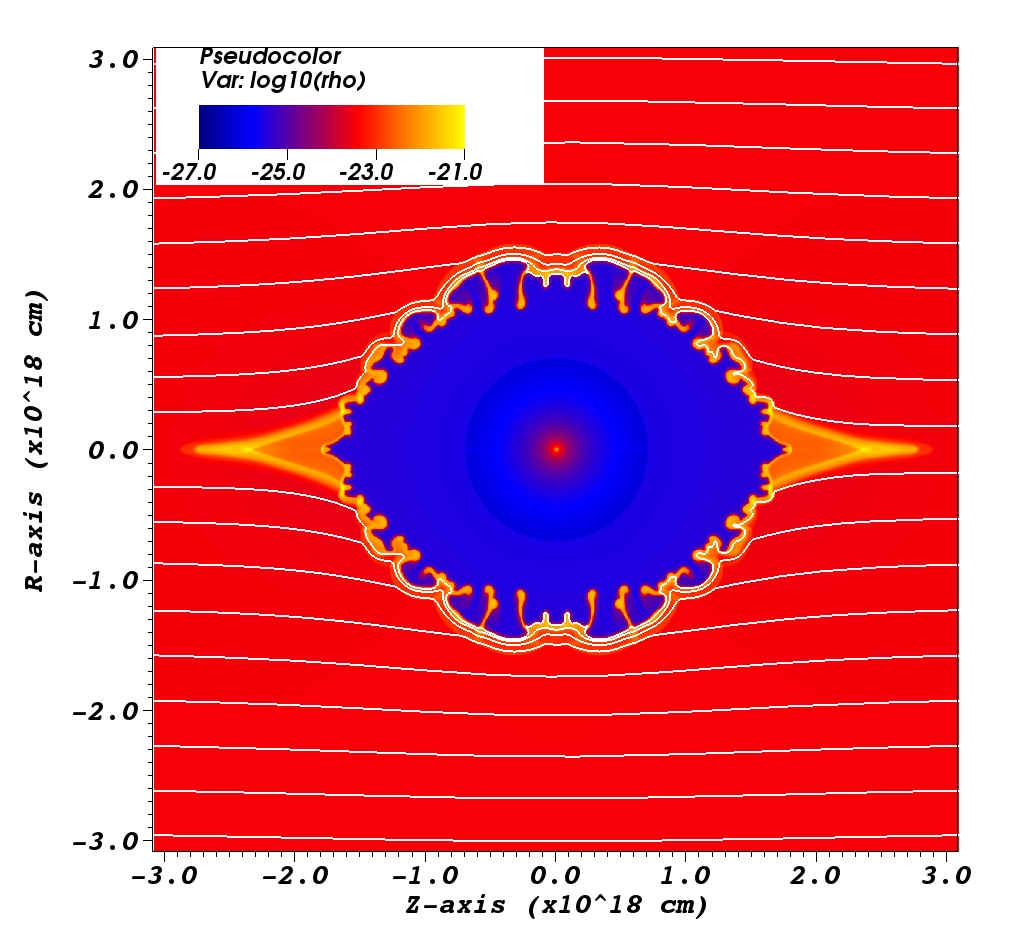}}}
}
\caption{Similar to Fig.~\ref{fig:PNE1}, but 6\,000 (left) and 8\,000 years (right) after the transition from AGB to post-AGB. 
The PN becomes ovoid due to the asymmetry of the medium into which it expands. 
Note the dependence of the wavelength of the instabilities on the point at which the shell interacts with the magnetic field (right).}
 \label{fig:PNE2}
\end{figure*}

\begin{figure*}
\FIG{
 \centering
\mbox{
\subfigure
{\includegraphics[width=0.5\textwidth]{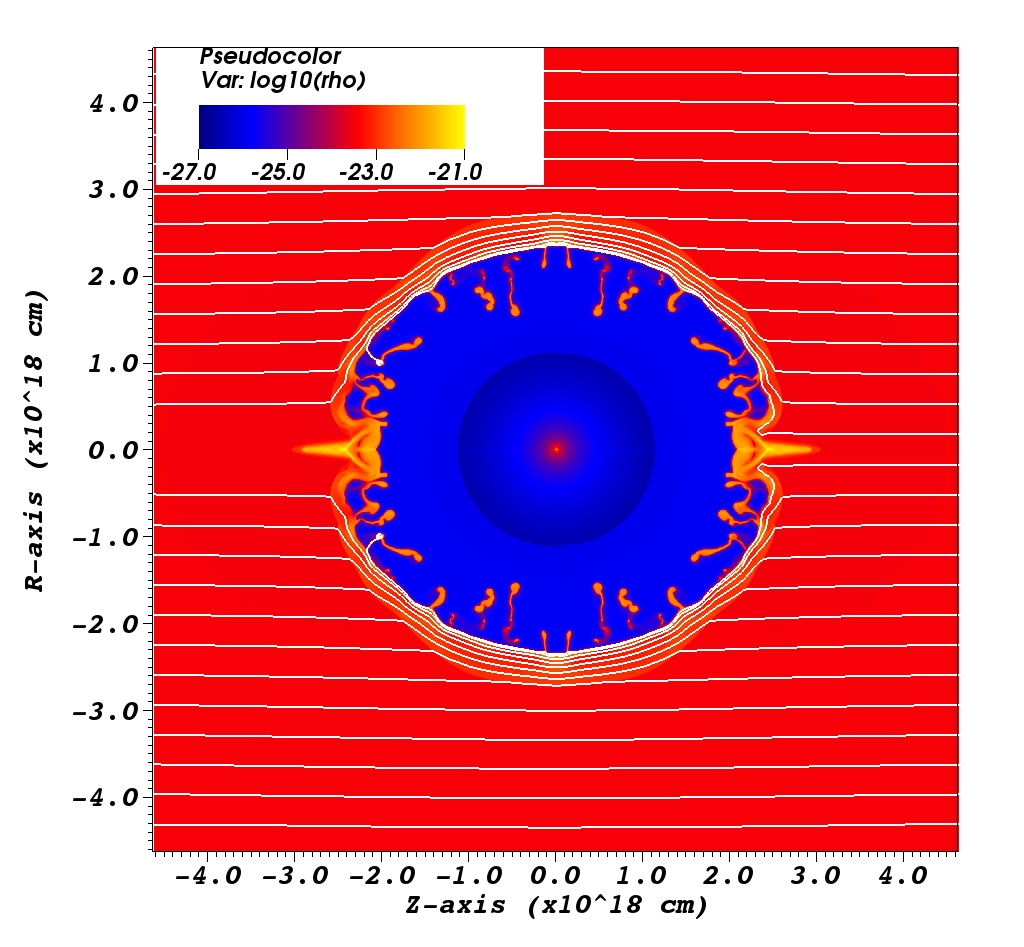}}
\subfigure
{\includegraphics[width=0.5\textwidth]{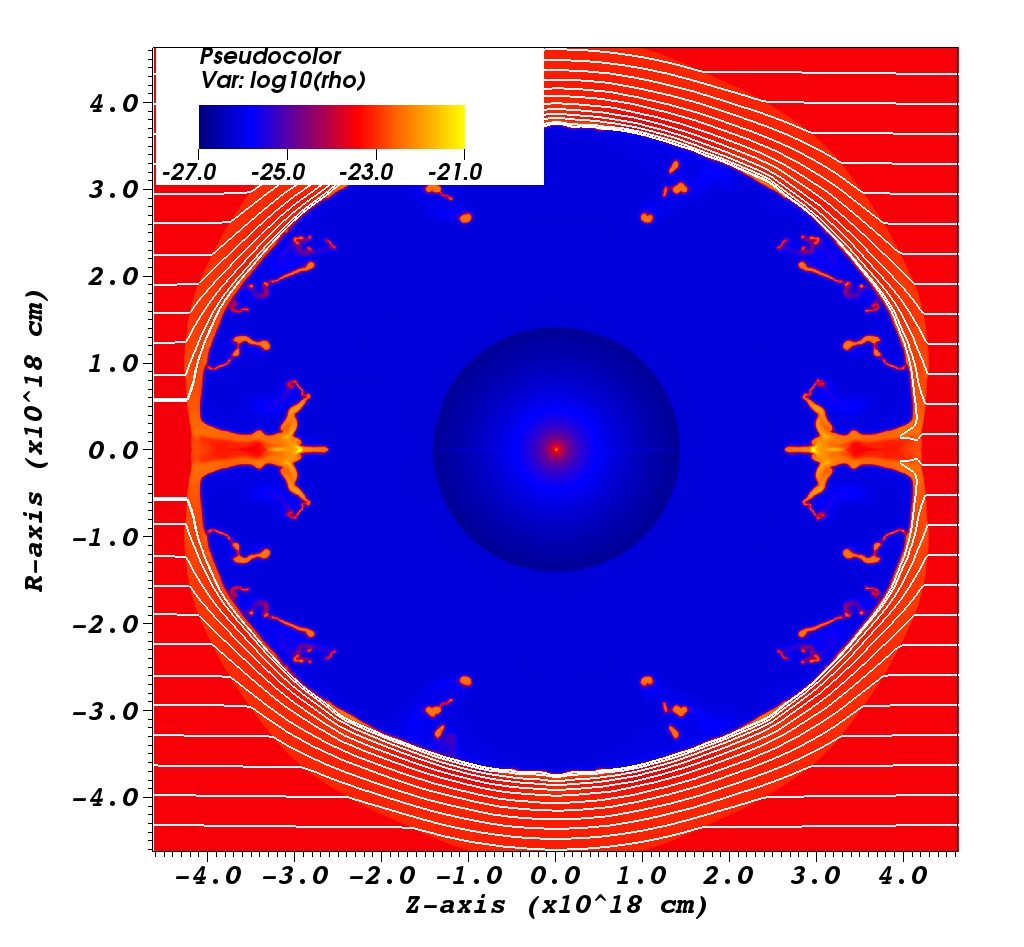}}}
}
\caption{Similar to Figs.~\ref{fig:PNE1}-\ref{fig:PNE2}, but 15\,000 (left) and 30\,000 years (right) after the transition from AGB to post-AGB. 
The PN encounters the high density points of the \emph{eye}, which slows down the expansion along the z-axis.}
 \label{fig:PNE3}
\end{figure*}

\section{Conclusions}
Even a weak interstellar magnetic field can distort an AGB wind bubble into an \emph{eye}-like shape. 
However, whether such a shape would actually be observed depends on other parameters, such as the stellar motion through the ISM 
as well as on the visibility of the shell at the time that it has an \emph{eye}-like shape. 
From our models we can conclude that \emph{visible} \emph{eye}-like nebulae are most likely to occur around stars that 
are subject to one or more of the following conditions:
\begin{enumerate}
 \item Stars that are situated in a (relatively) strong interstellar magnetic field ($\simgr5$\muG). 
 This reduces the distance from the star at which the magnetic tension force starts to influence the shape of the shell. 
 \item Stars that are situated in a low density ISM ($\simle2\,/{\rm cm}^3$). 
 This reduces the influence of stellar motion and allows the magnetic field to dominate the restraining force that acts on the outside 
 of the shell. Low density regions also tend to have a higher ISM temperature, which reduces the size of the nebula. This increases 
 the density in the shell as well as the dust temperature, making the shell more observable. 
 \item Stars that have a relatively weak wind, allowing the magnetic field to do dominate the force balance at an earlier stage. 
 \item Stars that have jet spent only a relatively short time in the AGB phase. Once the star grows older, the \emph{eye}-like structure disappears and the nebula 
 grows larger, making it more difficult to observe. 
 \item Stars that have low proper motion (at least compared to the local sound speed). 
 This gives the star more time to form an \emph{eye}-like nebula, 
 before the stellar motion dominates the shape of the shell to form a bow shock. 
\end{enumerate}

In the direction perpendicular to the magnetic field, the AGB wind bubble remains small. 
Depending on mass-loss rate, magnetic field strength, and ISM density, the final short-axis radius of the bubble varies between 0.1 and 0.6 pc. 
This has potential implications for the evolution of PNe, which may reach the outer edge of the AGB-wind bubble 
within their $\sim10\,000$\,year life\-span.

\begin{acknowledgements} 
We thank our anonymous referee for many insightful comments that helped us to improve our paper. 
A.J.v.M.\ acknowledges support from FWO, grant G.0277.08, KU~Leuven GOA/2008/04 and GOA/2009/09. 
N.L.J.C.\ and L.D.\ acknowledge support from the Belgian Federal Science Policy Office via the PRODEX Programme of ESA.
PACS has been developed by a consortium of institutes led by MPE (Germany) and including UVIE (Austria); 
KU Leuven, CSL, IMEC (Belgium); CEA, LAM (France); MPIA (Germany); INAF-IFSI/OAA/OAP/OAT, LENS, SISSA (Italy); IAC (Spain). 
This development has been supported by the funding agencies BMVIT (Austria), ESA-PRODEX (Belgium), CEA/CNES (France), 
DLR (Germany), ASI/INAF (Italy), and CICYT/MCYT (Spain).
\end{acknowledgements}

\bibliographystyle{aa}
\bibliography{vanmarle_biblio}

\begin{thebibliography}{64}
\expandafter\ifx\csname natexlab\endcsname\relax\def\natexlab#1{#1}\fi

\bibitem[{{Andersen}(2007)}]{Andersen:2007}
{Andersen}, A.~C. 2007, in Astronomical Society of the Pacific Conference
  Series, Vol. 378, Why Galaxies Care About AGB Stars: Their Importance as
  Actors and Probes, ed. F.~{Kerschbaum}, C.~{Charbonnel}, \& R.~F. {Wing}, 170

\bibitem[{{Arndt} {et~al.}(1997){Arndt}, {Fleischer}, \&
  {Sedlmayr}}]{Arndtetal:1997}
{Arndt}, T.~U., {Fleischer}, A.~J., \& {Sedlmayr}, E. 1997, \aap, 327, 614

\bibitem[{{Avedisova}(1972)}]{Avedisova:1972}
{Avedisova}, V.~S. 1972, \sovast, 15, 708

\bibitem[{{Bergeat} \& {Chevallier}(2005)}]{BergeatChevallier:2005}
{Bergeat}, J. \& {Chevallier}, L. 2005, \aap, 429, 235

\bibitem[{{Binney} \& {Merrifield}(1998)}]{BinneyMerrifield:1998}
{Binney}, J. \& {Merrifield}, M. 1998, {Galactic Astronomy}

\bibitem[{{Breitschwerdt} {et~al.}(2000){Breitschwerdt}, {Freyberg}, \&
  {Egger}}]{Breitschwerdtetal:2000}
{Breitschwerdt}, D., {Freyberg}, M.~J., \& {Egger}, R. 2000, A\&A, 361, 303

\bibitem[{{Caselli} {et~al.}(1998){Caselli}, {Walmsley}, {Terzieva}, \&
  {Herbst}}]{Casellietal:1998}
{Caselli}, P., {Walmsley}, C.~M., {Terzieva}, R., \& {Herbst}, E. 1998, \apj,
  499, 234

\bibitem[{{Clark} {et~al.}(2010){Clark}, {Garc{\'{\i}}a-D{\'{\i}}az},
  {L{\'o}pez}, {Steffen}, \& {Richer}}]{Clarketal:2010}
{Clark}, D.~M., {Garc{\'{\i}}a-D{\'{\i}}az}, M.~T., {L{\'o}pez}, J.~A.,
  {Steffen}, W.~G., \& {Richer}, M.~G. 2010, \apj, 722, 1260

\bibitem[{{Cox} {et~al.}(2012){Cox}, {Kerschbaum}, {van Marle}, {Decin},
  {Ladjal}, {Mayer}, {Groenewegen}, {van Eck}, {Royer}, {Ottensamer}, {Ueta},
  {Jorissen}, {Mecina}, {Meliani}, {Luntzer}, {Blommaert}, {Posch},
  {Vandenbussche}, \& {Waelkens}}]{Coxetal:2012}
{Cox}, N.~L.~J., {Kerschbaum}, F., {van Marle}, A.-J., {et~al.} 2012, \aap,
  537, A35

\bibitem[{{Dalgarno} \& {McCray}(1972)}]{DalgarnoMcCray:1972}
{Dalgarno}, A. \& {McCray}, R.~A. 1972, \araa, 10, 375

\bibitem[{{Decin} {et~al.}(2012){Decin}, {Cox}, {Royer}, {Van Marle},
  {Vandenbussche}, {Ladjal}, {Kerschbaum}, {Ottensamer}, {Barlow}, {Blommaert},
  {Gomez}, {Groenewegen}, {Lim}, {Swinyard}, {Waelkens}, \&
  {Tielens}}]{Decinetal:2012}
{Decin}, L., {Cox}, N.~L.~J., {Royer}, P., {et~al.} 2012, \aap, 548, A113

\bibitem[{{Elitzur} \& {Ivezi{\'c}}(2001)}]{ElitzurIvezi:2001}
{Elitzur}, M. \& {Ivezi{\'c}}, {\v Z}. 2001, \mnras, 327, 403

\bibitem[{{Falceta-Gon{\c c}alves} \&
  {Monteiro}(2014)}]{FalcetaGoncalvesMonteiro:2014}
{Falceta-Gon{\c c}alves}, D. \& {Monteiro}, H. 2014, \mnras

\bibitem[{{Fatuzzo} {et~al.}(2006){Fatuzzo}, {Adams}, \&
  {Melia}}]{Fatuzzoetal:2006}
{Fatuzzo}, M., {Adams}, F.~C., \& {Melia}, F. 2006, \apjl, 653, L49

\bibitem[{{Ferriere} {et~al.}(1991){Ferriere}, {Mac Low}, \&
  {Zweibel}}]{Ferriere:1991}
{Ferriere}, K.~M., {Mac Low}, M.-M., \& {Zweibel}, E.~G. 1991, \apj, 375, 239

\bibitem[{{Garc{\'{\i}}a-Segura} {et~al.}(1999){Garc{\'{\i}}a-Segura},
  {Langer}, {R{\'o}{\.z}yczka}, \& {Franco}}]{Garcia-Seguraetal:1999}
{Garc{\'{\i}}a-Segura}, G., {Langer}, N., {R{\'o}{\.z}yczka}, M., \& {Franco},
  J. 1999, \apj, 517, 767

\bibitem[{{Groenewegen} {et~al.}(2011){Groenewegen}, {Waelkens}, {Barlow},
  {Kerschbaum}, {Garcia-Lario}, {Cernicharo}, {Blommaert}, {Bouwman}, {Cohen},
  {Cox}, {Decin}, {Exter}, {Gear}, {Gomez}, {Hargrave}, {Henning},
  {Hutsem{\'e}kers}, {Ivison}, {Jorissen}, {Krause}, {Ladjal}, {Leeks}, {Lim},
  {Matsuura}, {Naz{\'e}}, {Olofsson}, {Ottensamer}, {Polehampton}, {Posch},
  {Rauw}, {Royer}, {Sibthorpe}, {Swinyard}, {Ueta}, {Vamvatira-Nakou},
  {Vandenbussche}, {van de Steene}, {van Eck}, {van Hoof}, {van Winckel},
  {Verdugo}, \& {Wesson}}]{Groenewegenetal:2011}
{Groenewegen}, M.~A.~T., {Waelkens}, C., {Barlow}, M.~J., {et~al.} 2011, \aap,
  526, A162

\bibitem[{{Guerrero} {et~al.}(2012){Guerrero}, {Ruiz}, {Hamann}, {Chu}, {Todt},
  {Sch{\"o}nberner}, {Oskinova}, {Gruendl}, {Steffen}, {Blair}, \&
  {Toal{\'a}}}]{Guerreroetal:2012}
{Guerrero}, M.~A., {Ruiz}, N., {Hamann}, W.-R., {et~al.} 2012, \apj, 755, 129

\bibitem[{{Habing} \& {Olofsson}(2003)}]{HabingOlofsson:2003}
{Habing}, H.~J. \& {Olofsson}, H., eds. 2003, {Asymptotic giant branch stars}

\bibitem[{{Han} {et~al.}(2004){Han}, {Ferriere}, \&
  {Manchester}}]{Hanetal:2004}
{Han}, J.~L., {Ferriere}, K., \& {Manchester}, R.~N. 2004, \apj, 610, 820

\bibitem[{{Heiligman}(1980)}]{Heiligman:1980}
{Heiligman}, G.~M. 1980, \mnras, 191, 761

\bibitem[{{Herwig}(2005)}]{Herwig:2005}
{Herwig}, F. 2005, \araa, 43, 435

\bibitem[{{Hoogzaad} {et~al.}(2002){Hoogzaad}, {Molster}, {Dominik}, {Waters},
  {Barlow}, \& {de Koter}}]{Hoogzaadetal:2002}
{Hoogzaad}, S.~N., {Molster}, F.~J., {Dominik}, C., {et~al.} 2002, \aap, 389,
  547

\bibitem[{{Jun} {et~al.}(1995){Jun}, {Norman}, \& {Stone}}]{Junetal:1995}
{Jun}, B.-I., {Norman}, M.~L., \& {Stone}, J.~M. 1995, ApJ, 453, 332

\bibitem[{{Kaastra} \& {Mewe}(2000)}]{KaastraMewe:2000}
{Kaastra}, J.~S. \& {Mewe}, R. 2000, in Atomic Data Needs for X-ray Astronomy,
  ed. M.~A. {Bautista}, T.~R. {Kallman}, \& A.~K. {Pradhan}, 161

\bibitem[{{Keppens} {et~al.}(2012){Keppens}, {Meliani}, {van Marle}, {Delmont},
  {Vlasis}, \& {van der Holst}}]{Keppensetal:2012}
{Keppens}, R., {Meliani}, Z., {van Marle}, A.~J., {et~al.} 2012, Journal of
  Computational Physics, 231, 718

\bibitem[{{Kerschbaum} {et~al.}(2010){Kerschbaum}, {Ladjal}, {Ottensamer},
  {Groenewegen}, {Mecina}, {Blommaert}, {Baumann}, {Decin}, {Vandenbussche},
  {Waelkens}, {Posch}, {Huygen}, {De Meester}, {Regibo}, {Royer}, {Exter}, \&
  {Jean}}]{Kerschbaumetal:2010}
{Kerschbaum}, F., {Ladjal}, D., {Ottensamer}, R., {et~al.} 2010, \aap, 518,
  L140

\bibitem[{{Kwok}(1975)}]{Kwok:1975}
{Kwok}, S. 1975, \apj, 198, 583

\bibitem[{{Kwok}(2000)}]{Kwok:2000}
{Kwok}, S. 2000, {The Origin and Evolution of Planetary Nebulae}, ed. {Kwok,
  S.}

\bibitem[{{Kwok} {et~al.}(1978){Kwok}, {Purton}, \&
  {Fitzgerald}}]{Kwoketal:1978}
{Kwok}, S., {Purton}, C.~R., \& {Fitzgerald}, P.~M. 1978, \apjl, 219, L125

\bibitem[{{Lamers} \& {Cassinelli}(1999)}]{Lamerscassinelli:1999}
{Lamers}, H.~J.~G.~L.~M. \& {Cassinelli}, J.~P. 1999, {Introduction to Stellar
  Winds} ({Cambridge, UK: Cambridge University Press})

\bibitem[{{Maercker} {et~al.}(2010){Maercker}, {Olofsson}, {Eriksson},
  {Gustafsson}, \& {Sch{\"o}ier}}]{Maerckeretal:2005}
{Maercker}, M., {Olofsson}, H., {Eriksson}, K., {Gustafsson}, B., \&
  {Sch{\"o}ier}, F.~L. 2010, \aap, 511, A37

\bibitem[{{Marengo} {et~al.}(1997){Marengo}, {Canil}, {Silvestro}, {Origlia},
  {Busso}, \& {Persi}}]{Marengoetal:1997}
{Marengo}, M., {Canil}, G., {Silvestro}, G., {et~al.} 1997, \aap, 322, 924

\bibitem[{{Mayer} {et~al.}(2013){Mayer}, {Jorissen}, {Kerschbaum},
  {Ottensamer}, {Nowotny}, {Cox}, {Aringer}, {Blommaert}, {Decin}, {van Eck},
  {Gail}, {Groenewegen}, {Kornfeld}, {Mecina}, {Posch}, {Vandenbussche}, \&
  {Waelkens}}]{Mayeretal:2013}
{Mayer}, A., {Jorissen}, A., {Kerschbaum}, F., {et~al.} 2013, \aap, 549, A69

\bibitem[{{Minter} \& {Spangler}(1996)}]{MinterSpangler:1996}
{Minter}, A.~H. \& {Spangler}, S.~R. 1996, \apj, 458, 194

\bibitem[{{Miszalski} {et~al.}(2012){Miszalski}, {Boffin}, {Frew}, {Acker},
  {K{\"o}ppen}, {Moffat}, \& {Parker}}]{Miszalskietal:2012}
{Miszalski}, B., {Boffin}, H.~M.~J., {Frew}, D.~J., {et~al.} 2012, \mnras, 419,
  39

\bibitem[{{Pilbratt} {et~al.}(2010){Pilbratt}, {Riedinger}, {Passvogel},
  {Crone}, {Doyle}, {Gageur}, {Heras}, {Jewell}, {Metcalfe}, {Ott}, \&
  {Schmidt}}]{Pilbrattetal:2010}
{Pilbratt}, G.~L., {Riedinger}, J.~R., {Passvogel}, T., {et~al.} 2010, \aap,
  518, L1

\bibitem[{{Poglitsch} {et~al.}(2010){Poglitsch}, {Waelkens}, {Geis},
  {Feuchtgruber}, {Vandenbussche}, {Rodriguez}, {Krause}, {Renotte}, {van
  Hoof}, {Saraceno}, {Cepa}, {Kerschbaum}, {Agn{\`e}se}, {Ali}, {Altieri},
  {Andreani}, {Augueres}, {Balog}, {Barl}, {Bauer}, {Belbachir}, {Benedettini},
  {Billot}, {Boulade}, {Bischof}, {Blommaert}, {Callut}, {Cara}, {Cerulli},
  {Cesarsky}, {Contursi}, {Creten}, {De Meester}, {Doublier}, {Doumayrou},
  {Duband}, {Exter}, {Genzel}, {Gillis}, {Gr{\"o}zinger}, {Henning},
  {Herreros}, {Huygen}, {Inguscio}, {Jakob}, {Jamar}, {Jean}, {de Jong},
  {Katterloher}, {Kiss}, {Klaas}, {Lemke}, {Lutz}, {Madden}, {Marquet},
  {Martignac}, {Mazy}, {Merken}, {Montfort}, {Morbidelli}, {M{\"u}ller},
  {Nielbock}, {Okumura}, {Orfei}, {Ottensamer}, {Pezzuto}, {Popesso},
  {Putzeys}, {Regibo}, {Reveret}, {Royer}, {Sauvage}, {Schreiber}, {Stegmaier},
  {Schmitt}, {Schubert}, {Sturm}, {Thiel}, {Tofani}, {Vavrek}, {Wetzstein},
  {Wieprecht}, \& {Wiezorrek}}]{Poglitschetal:2010}
{Poglitsch}, A., {Waelkens}, C., {Geis}, N., {et~al.} 2010, \aap, 518, L2

\bibitem[{{Powell} {et~al.}(1999){Powell}, {Roe}, {Linde}, {Gombosi}, \& {de
  Zeeuw}}]{Powelletal:1999}
{Powell}, K.~G., {Roe}, P.~L., {Linde}, T.~J., {Gombosi}, T.~I., \& {de Zeeuw},
  D.~L. 1999, Journal of Computational Physics, 154, 284

\bibitem[{{Ransom} {et~al.}(2008){Ransom}, {Uyaniker}, {Kothes}, \&
  {Landecker}}]{Ransometal:2008}
{Ransom}, R.~R., {Uyaniker}, B., {Kothes}, R., \& {Landecker}, T.~L. 2008,
  \apj, 684, 1009

\bibitem[{{Sch{\"o}nrich} {et~al.}(2010){Sch{\"o}nrich}, {Binney}, \&
  {Dehnen}}]{Schronrichetal:2010}
{Sch{\"o}nrich}, R., {Binney}, J., \& {Dehnen}, W. 2010, \mnras, 403, 1829

\bibitem[{{Schure} {et~al.}(2009){Schure}, {Kosenko}, {Kaastra}, {Keppens}, \&
  {Vink}}]{Schureetal:2009}
{Schure}, K.~M., {Kosenko}, D., {Kaastra}, J.~S., {Keppens}, R., \& {Vink}, J.
  2009, \aap, 508, 751

\bibitem[{{Shu}(1983)}]{Shu:1983}
{Shu}, F.~H. 1983, \apj, 273, 202

\bibitem[{{Soker} \& {Dgani}(1997)}]{SokerDgani:1997}
{Soker}, N. \& {Dgani}, R. 1997, \apj, 484, 277

\bibitem[{{Soker} \& {Zucker}(1997)}]{SokerZucker:1997}
{Soker}, N. \& {Zucker}, D.~B. 1997, \mnras, 289, 665

\bibitem[{{Stevens} {et~al.}(1992){Stevens}, {Blondin}, \&
  {Pollock}}]{Stevensetal:1992}
{Stevens}, I.~R., {Blondin}, J.~M., \& {Pollock}, A.~M.~T. 1992, \apj, 386, 265

\bibitem[{{Tanab{\'e}} {et~al.}(1997){Tanab{\'e}}, {Nishida}, {Matsumoto},
  {Onaka}, {Nakada}, {Soyano}, {Ono}, {Sekiguchi}, \&
  {Glass}}]{Tanabeetal:1997}
{Tanab{\'e}}, T., {Nishida}, S., {Matsumoto}, S., {et~al.} 1997, \nat, 385, 509

\bibitem[{{Terzian}(1997)}]{Terzian:1997}
{Terzian}, Y. 1997, in IAU Symposium, Vol. 180, Planetary Nebulae, ed. H.~J.
  {Habing} \& H.~J.~G.~L.~M. {Lamers}, 29

\bibitem[{{Tielens}(2005)}]{Tielens:2005}
{Tielens}, A.~G.~G.~M. 2005, {The Physics and Chemistry of the Interstellar
  Medium}

\bibitem[{{Tomisaka}(1990)}]{Tomisaka:1990}
{Tomisaka}, K. 1990, \apjl, 361, L5

\bibitem[{{Tomisaka}(1992)}]{Tomisaka:1992}
{Tomisaka}, K. 1992, \pasj, 44, 177

\bibitem[{{Tweedy} {et~al.}(1995){Tweedy}, {Martos}, \&
  {Noriega-Crespo}}]{Tweedyetal:1995}
{Tweedy}, R.~W., {Martos}, M.~A., \& {Noriega-Crespo}, A. 1995, \apj, 447, 257

\bibitem[{{van Marle} {et~al.}(2014){van Marle}, {Decin}, \&
  {Meliani}}]{vanMarleetal:2014}
{van Marle}, A.~J., {Decin}, L., \& {Meliani}, Z. 2014, \aap, 561, A152

\bibitem[{{van Marle} {et~al.}(2011){van Marle}, {Meliani}, {Keppens}, \&
  {Decin}}]{vanMarleetaldust:2011}
{van Marle}, A.~J., {Meliani}, Z., {Keppens}, R., \& {Decin}, L. 2011, \apjl,
  734, L26+

\bibitem[{{Villaver} {et~al.}(2002){Villaver}, {Garc{\'{\i}}a-Segura}, \&
  {Manchado}}]{Villaveretal:2002}
{Villaver}, E., {Garc{\'{\i}}a-Segura}, G., \& {Manchado}, A. 2002, \apj, 571,
  880

\bibitem[{{Villaver} {et~al.}(2003){Villaver}, {Garc{\'{\i}}a-Segura}, \&
  {Manchado}}]{Villaveretal:2003}
{Villaver}, E., {Garc{\'{\i}}a-Segura}, G., \& {Manchado}, A. 2003, \apjl, 585,
  L49

\bibitem[{{Villaver} {et~al.}(2012){Villaver}, {Manchado}, \&
  {Garc{\'{\i}}a-Segura}}]{Villaveretal:2012}
{Villaver}, E., {Manchado}, A., \& {Garc{\'{\i}}a-Segura}, G. 2012, \apj, 748,
  94

\bibitem[{{Vishniac}(1983)}]{Vishniac:1983}
{Vishniac}, E.~T. 1983, \apj, 274, 152

\bibitem[{{Wareing} {et~al.}(2007){Wareing}, {Zijlstra}, \&
  {O'Brien}}]{Wareingetal:2007}
{Wareing}, C.~J., {Zijlstra}, A.~A., \& {O'Brien}, T.~J. 2007, \mnras, 382,
  1233

\bibitem[{{Weaver} {et~al.}(1977){Weaver}, {McCray}, {Castor}, {Shapiro}, \&
  {Moore}}]{Weaveretal:1977}
{Weaver}, R., {McCray}, R., {Castor}, J., {Shapiro}, P., \& {Moore}, R. 1977,
  \apj, 218, 377

\bibitem[{{Wilkin}(1996)}]{Wilkin:1996}
{Wilkin}, F.~P. 1996, \apjl, 459, L31+

\bibitem[{{Woods} {et~al.}(2012){Woods}, {Walsh}, {Cordiner}, \&
  {Kemper}}]{Woodsetal:2012}
{Woods}, P.~M., {Walsh}, C., {Cordiner}, M.~A., \& {Kemper}, F. 2012, \mnras,
  426, 2689

\bibitem[{{Zhukovska} {et~al.}(2008){Zhukovska}, {Gail}, \&
  {Trieloff}}]{Shukovskaetal:2008}
{Zhukovska}, S., {Gail}, H.-P., \& {Trieloff}, M. 2008, \aap, 479, 453

\bibitem[{{Zhukovska} \& {Henning}(2013)}]{ZhukovskaHenning:2013}
{Zhukovska}, S. \& {Henning}, T. 2013, \aap, 555, A99

\end{thebibliography}

\IfFileExists{vanmarle_biblio.bbl}{}
 {\typeout{}
  \typeout{******************************************}
  \typeout{** Please run "bibtex \jobname" to obtain}
  \typeout{** the bibliography and then re-run LaTeX}
  \typeout{** twice to fix the references!}
  \typeout{******************************************}
  \typeout{}
 }
 
\listofobjects
\Online
\begin{appendix}
\section{Animations}
\begin{figure}
\caption{Animation of Simulation~A1}
\end{figure}

\begin{figure}
\caption{Animation of Simulation~A2} 
\end{figure}

\begin{figure}
\caption{Animation of Simulation~B1}
\end{figure}

\begin{figure}
\caption{Animation of Simulation~B2} 
\end{figure}

\begin{figure}
\caption{Animation of Simulation~B3}
\end{figure}

\begin{figure}
\caption{Animation of Simulation~B4}
\end{figure}

\begin{figure}
\caption{Animation of Simulation~C1}
\end{figure}

\begin{figure}
\caption{Animation of Simulation~C2} 
\end{figure}

\begin{figure}
\caption{Animation of Simulation~C3}
\end{figure}

\begin{figure}
\caption{Animation of Simulation~C4}
\end{figure}
 
\begin{figure}
\caption{Animation of Simulation~E1}
\end{figure}
  
\begin{figure}
\caption{Animation of Planetary Nebula}
\end{figure}

\end{appendix}

\end{document}